\documentclass[prl,aps,amsmath,amssymb,reprint]{revtex4-2}
\usepackage{color}
\usepackage{amsmath}
\usepackage{bm}
\usepackage{multirow}
\usepackage{graphicx}
\usepackage{amsthm}

\begin{document}
\title{Heterogeneous nucleation in finite size adaptive dynamical networks} 
	


\author{Jan Fialkowski$^{1}$}
\email[]{jan.fialkowski@campus.tu-berlin.de}
\author{Serhiy Yanchuk$^{2,3}$}
\author{Igor M. Sokolov$^{4}$}
\author{Eckehard Sch\"oll$^{1,2,5}$}
\author{Georg A. Gottwald$^{6}$}
\author{Rico Berner$^{1,4}$}
\email[]{rico.berner@physik.hu-berlin.de}
\affiliation{$^{1}$Institute for Theoretical Physics, Technische Universit\"at Berlin, Hardenbergstr.\,36, 10623 Berlin, Germany}
\affiliation{$^{2}$Potsdam Institute for Climate Impact Research, Telegrafenberg A 31, 14473 Potsdam, Germany}
\affiliation{$^{3}$Institute of Mathematics, Humboldt University Berlin, 12489 Berlin, Germany}
\affiliation{$^{4}$Department of Physics, Humboldt Universit\"at zu Berlin, Newtonstraße 15, 12489 Berlin, Germany}
\affiliation{$^{5}$Bernstein Center for Computational Neuroscience Berlin, Humboldt Universit\"at, 10115 Berlin, Germany}
\affiliation{$^{6}$School of Mathematics and Statistics, University of Sydney, Camperdown NSW 2006, Australia}


\date{\today}
\begin{abstract}
    Phase transitions in equilibrium and nonequilibrium systems play a major role in the natural sciences. In dynamical networks, phase transitions organize qualitative changes in the collective behavior of coupled dynamical units. Adaptive dynamical networks feature a connectivity structure that changes over time, co-evolving with the nodes' dynamical state. 
    In this Letter, we show the emergence of two distinct first-order nonequilibrium phase transitions in a finite size adaptive network of heterogeneous phase oscillators. Depending on the nature of defects in the internal frequency distribution, we observe either an abrupt single-step transition to full synchronization or a more gradual multi-step transition. This observation has a striking resemblance to heterogeneous nucleation. We develop a mean-field approach to study the interplay between adaptivity and nodal heterogeneity and describe the dynamics of multicluster states and their role in determining the character of the phase transition. Our work provides a theoretical framework for studying the interplay between adaptivity and nodal heterogeneity.
\end{abstract} 
\maketitle

Phase transitions play an important role in the natural sciences~\cite{STA71}. 
Complex dynamical networks~\cite{NEW03,BOC18} exhibit a plethora of nonequilibrium phase transitions organizing their collective dynamics in response to variations in control parameters such as interaction strength or noise~\cite{HAK83,SCH87}. In particular, transitions between coherence and incoherence have attracted significant attention in static~\cite{ROD16} and temporally evolving complex networks~\cite{GHO22}. 
The Kuramoto model~\cite{KUR84} has served as a testbed to study phase transitions in networks of coupled oscillators. It exhibits either first or second-order phase transitions from incoherence to full synchronization, depending on the natural frequency distribution~\cite{KUR84,PIK01,PAZ05a,GOM11a,BOC16,SOU19}. Similarly, network structure~\cite{GOM07} and weight distribution~\cite{ZHA13a} can lead to first-order (or explosive) transitions and to hysteresis, for which a universal mechanism has recently been proposed~\cite{KUE21}.

To better describe real-world phenomena, the original Kuramoto model has been extended and modified. Beyond the classical Kuramoto model, generalizations to static and time-evolving networks have been developed
~\cite{YEU99,STR00,LEE09,BRE10h,NAB11,BIC20,GHO22}. 
The inclusion of additional dynamical degrees of freedom, e.g. to describe power grids~\cite{OLM14a,TUM18,TUM19,TAH19,HEL20} by including inertia, has introduced much richer synchronization transitions with regimes of coexisting cluster states.
Recently, adaptive dynamical network models were introduced which are capable of describing chemical~\cite{JAI01,KUE19a}, epidemic~\cite{GRO06b}, biological~\cite{PRO05a}, neurological~\cite{GER96,MEI09a,MIK13}, transport~\cite{MAR17b}, and social systems~\cite{GRO08a,HOR20}. Adaptive dynamical networks are characterized by the coevolution of network structure and functionality. Paradigmatic models of adaptively coupled phase oscillators have recently attracted much attention~\cite{GUT11,KAS17,BER19,BER20,FEK20,BER21b}. They have shown promise in predicting and describing phenomena in more realistic and complex physical systems such as neuronal and biological systems~\cite{LUE16,ROE19a,SAW21b,BER22}, as well as power grid models~\cite{BER21a}.
However, the type and nature of phase transition in this important class of models remains unclear. 

This work characterizes nonequilibrium phase transitions in adaptive networks. We describe phase transitions in a finite-size Kuramoto model equipped with adaptive coupling weights. The natural frequencies are considered to be uniformly distributed, i.e., all possible frequencies are equally probable, and therefore disorders induced by finite size realizations of this distribution impact directly the synchronization behavior. We find two qualitatively distinct types of first-order transitions to synchrony akin to first-order transition phenomena of heterogeneous nucleation~\cite{THA14}.
The first, \textit{multi-step} type of synchronization transition is characterized by the nucleation and growth of a dominant cluster, similar to Ostwald ripening in equilibrium and nonequilibrium systems~\cite{SCH91}, until the system reaches synchrony. The second, \textit{single-step} transition type features multiple stable synchronization nuclei and the transitions to full synchrony is caused by an abrupt merging of large clusters of similar size. These two paths to synchrony exhibit a high degree of multistability. We identify the location of fluctuations in the realization of the natural frequency distribution as the cause for the two different scenarios.    
Methodologically, we present a framework reducing high dimensional adaptive networks to a few mesoscopic variables and show that a collection of partially synchronized clusters of approximately equal size is more stable against changes in the coupling strength. With this, we extend the scope of mean-field approaches beyond the static network paradigm to adaptive networks.


For neuronal systems with spike-timing-dependent synaptic plasticity, phase oscillator models with phase difference-dependent adaptation functions have been introduced to explain effects stemming from long-term potentiation and depression~\cite{AOK09,MUL11,LUE16,KAS17,ROE19a}. Beyond that, phase oscillator models have served as paradigms for studying collective behavior in real-world dynamical systems~\cite{ACE05}. We consider the adaptively coupled phase oscillator model 
\begin{align}
	\frac{d\phi_i}{dt}&=\omega_i-\frac{\sigma}{N}\sum_{j=1}^N\kappa_{ij}\sin(\phi_i-\phi_j)\label{EQ:Phidot},\\
	\frac{d\kappa_{ij}}{dt}&=-\epsilon\left(\kappa_{ij}+\sin(\phi_i-\phi_j+\beta)\right),\label{EQ:Kappadot}
\end{align}
for $N$ phase oscillators $i=1,...,N$ with phases $\phi_i(t)\in [0,2\pi)$ coevolving with adaptive coupling weights $\kappa_{ij}(t)$. The natural frequency $\omega_i$ of the $i$th oscillator is drawn randomly from a uniform distribution $\omega_i\in \left[-\hat{\omega},~\hat{\omega}\right]$ \cite{Note1}. The overall coupling strength is $\sigma$, and $\epsilon$ characterizes the time-scale separation between the fast oscillator dynamics and the slower adaptation of the coupling weights. The parameter $\beta$ accounts for different adaptation rules, e.g., a causal rule ($\beta=0$) or a symmetric rule ($\beta=-\pi/2$) where the order of the phases or their closeness determines the sign of the coupling weight change~\cite{AOK09,BER19a}. Here we focus on symmetric adaptation rules which support synchronization~\cite{KAS17}; see~\cite{suppl} for other adaptation rules.

Coherence can be quantified by the synchronization index $S$ which measures the fraction of frequency-synchronized oscillator pairs $S:=\frac{1}{N^2}\sum_{i,j=1}^N s_{ij}$, where $s_{ij}=1$ for equal mean phase velocities of the oscillators $i$ and $j$, $\langle\dot{\phi}_i\rangle = \langle\dot{\phi}_j\rangle $, and $s_{ij}=0$ otherwise~\cite{Note2}. 
Here $\langle x\rangle=\lim_{T\to\infty}\frac{1}{T}~\int_{T_0}^{T_0+T}x(t)\mathrm{d}t$ with sufficiently large transient time $T_0$. For $S=1$ the system is fully frequency-synchronized, whereas for $S=0$ the system is asynchronous.

\begin{figure}[t]
    \centering
    \includegraphics[width=\columnwidth]{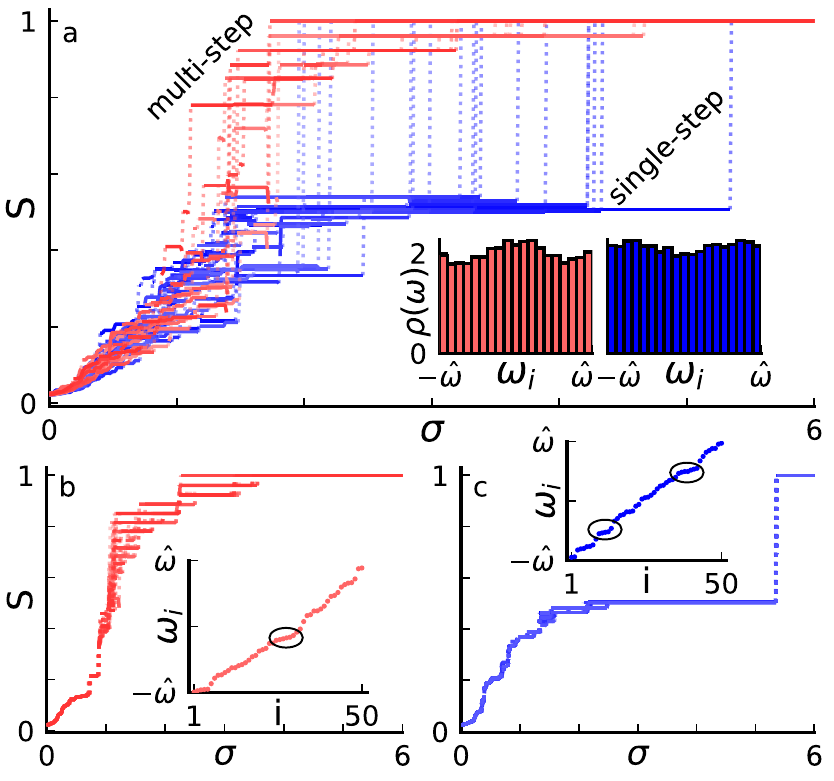}
    \caption{Paths to synchrony for system~\eqref{EQ:Phidot}--\eqref{EQ:Kappadot}. (a) Synchronization index $S$ as a function of the coupling strength $\sigma$ 
    for $100$ simulations with $N=50$ oscillators. Each run was initiated with random initial conditions and $\sigma$ was increased in steps of $\Delta\sigma=0.01$. For details on the up-sweep protocol, we refer to~\cite{suppl}. For each run, natural frequencies are drawn independently from a uniform distribution $\omega_i\in\left[\hat{\omega},~\hat{\omega}\right]$ with $\hat{\omega}=0.25$. The dotted lines indicate jumps in the synchronization index during the transition. The inset shows realizations of the finite-size frequency distributions for the multi-step (left) and single-step (right) paths generated from $1000$ simulations. (b) and (c): Synchronization index $S$ for fixed realization of the natural frequencies and $30$ different initial conditions, consistently leading to the multi-step (b) or single-step transition (c). Insets show the natural frequencies used in the simulations. The circles highlight areas of higher frequency densities. The synchronization index $S$ is determined with an averaging time window $T=7\cdot10^3$ and transient time $T_0=3\cdot10^3$. Other parameters: $\beta=-0.53\pi$, $\epsilon=0.01$.}
    \label{fig:Figure1}
\end{figure}

With increasing coupling strength $\sigma$, system~\eqref{EQ:Phidot}--\eqref{EQ:Kappadot} undergoes a transition from asynchrony to full synchronization, see Fig.~\ref{fig:Figure1}(a) and Supplemental Material~\cite{suppl}. 
The routes to synchrony with increasing coupling strength $\sigma$ in Fig.~\ref{fig:Figure1}(a) follow two paths of first-order transitions: a gradual multi-step  (upper, red) and an abrupt single-step (lower, blue) path featuring multiple small steps or a single large step in the transition to synchrony, respectively. 
In what follows, we describe these paths and determine the finite size features in the realization of the natural frequency distribution that lead to either a multi-step or single-step transition to synchrony.

\begin{figure}[t]
    \centering
    \includegraphics[width=\columnwidth]{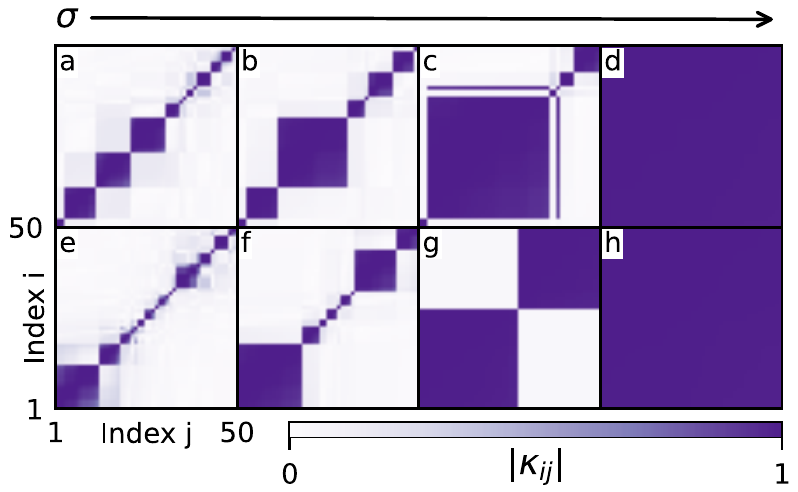}
    \caption{Coupling matrices for specific values of $\sigma$, corresponding to two different types of synchronization transition.
    (a)--(d): multi-step, and (e)--(h) single-step transition. 
    The values of $\sigma$ are (a) $0.9$, (b) $1.05$, (c) $1.75$, (d) $6.0$ and (e) $0.55$, (f) $1.75$ (g) $2.75$ and (h) $6.0$. 
    The snapshots are taken after $10^4$ time units. Other parameters: $\hat{\omega}=0.25$, $\beta=-0.53\pi$ and $\epsilon=0.01$. Due to the $\phi\rightarrow\phi+\pi$ symmetry, $\kappa_{ij}$ and $-\kappa_{ij}$ are indistinguishable, therefore the absolute values $|\kappa_{ij}|$ are plotted, see~\cite{suppl}.}
    \label{fig:Figure2}
\end{figure}

Figures~\ref{fig:Figure1}(b) and \ref{fig:Figure1}(c) show the multi-step and single-step transitions, respectively, for two representative realizations of the natural frequencies (displayed as insets) and $30$ different initial conditions of the phases $\phi_i(0)$. 
Depending on initial conditions and natural frequencies the system can develop a large number of coexisting states. The multi-step path in Fig.~\ref{fig:Figure1}(b) exhibits a higher degree of multistability even for the fixed realization of frequencies and a large number of transitions between coexisting states.
Figs.~\ref{fig:Figure2}(a)--(d) show snapshots of the coupling matrix corresponding to the multi-step transition. It is seen that a cluster nucleus emerges and, upon increasing the coupling strength, entrains more and more oscillators leading to the fully frequency synchronized state in Fig.~\ref{fig:Figure2}(d), similarly as observed in~\cite{OLM14a}.  
Depending on the initial conditions the exact structure of the system's state can vary, e.g. the size of the nucleus for different $\sigma$. This gives rise to a multitude of stable states for most values of $\sigma$. 

The single-step transition is shown in Fig.~\ref{fig:Figure1}(c). Regardless of the initial condition, the dynamics follows very similar paths of the synchronization index. The phase oscillators organize into a small number of clusters within each of which all oscillators move with the same mean phase velocity for small values of $\sigma$. Notably, there is no single cluster at any time that is significantly larger than the others, in contrast to the multi-step transition and prior findings for multicluster states~\cite{BER19}. 
Further, for the single-step phase transition path an intermediate state with $S\approx0.5$ emerges, which is stable for a wide range of coupling strengths $\sigma$. The transition to full frequency synchronization occurs discontinuously at $\sigma\approx 5.35$. The formation of small initial clusters and the intermediate state are shown in Figs.~\ref{fig:Figure2}(e)--(h). The intermediate state (g) consists of two almost evenly sized clusters (nuclei) that are formed simultaneously. 

Whether the system undergoes a multi-step or a single-step transition path is determined by the particular realization of the natural frequency distribution, see the insets in Fig~\ref{fig:Figure1}(a). Frequency distributions corresponding to multi-step phase transitions are characterized by a higher density around the average frequency $\bar \omega =0$, leading to an initial cluster with average cluster frequency close to the overall average frequency of the network. On the other hand, frequency distributions corresponding to single-step transitions are characterized by deviations which are concentrated away from the average frequency. This leads to local clusters emerging at low coupling strength around those seeds. These clusters survive a further increase in coupling strength entraining further oscillators before for larger coupling strength collapsing into the fully synchronized state.

The distinction between these two qualitatively very different scenarios of phase transitions is hence caused by finite-size-induced inhomogeneities 
of the natural frequency distribution. The fluctuations in the realization of the natural frequencies are much more influential than the choice of initial conditions, see Fig~\ref{fig:Figure1}(b) and (c).
To further probe that the choice of transition path is indeed a finite-size effect, we choose equidistantly distributed frequencies, to remove fluctuations. In this case the transition path which is taken by the system is determined instead by the initial condition of the phases, see~\cite{suppl}. 

Performing a down-sweep from large $\sigma$ for the given frequency distributions in Fig.~\ref{fig:Figure1}(b) and (c), hysteresis and bistability between full and partial cluster synchrony is observed, see~\cite{suppl}. 

During the transition to synchrony, oscillators group into phase-locked clusters. The fluctuations in the realization of the natural frequencies determine the shapes of the emerging cluster states which in turn govern the type of the transition. In systems with adaptive coupling weights there are many ways in which an arbitrary number of oscillators can form a multicluster structure~\cite{MAI07,BER19}. In the following, we study the coexistence of multicluster states and show that states with equally sized clusters are stable for a larger interval in $\sigma$ than states with strongly different cluster sizes. For this, we develop a mean-field description of multiclusters employing the collective coordinate method introduced in~\cite{GOT15}. The latter has been successfully used to describe the synchronization of the Kuramoto system for general frequency distributions~\cite{SMI19}, complex coupling topologies \cite{HAN18b} and chaotic cluster dynamics~\cite{SMI19}. This approach can capture finite-size effects while still reproducing the findings of mean-field theories~\cite{SMI20}. 

Following~\cite{SMI19}, we assign each oscillator to one of $M$ clusters $\mu$ of size $N_\mu=n_\mu N$. The phase variables and coupling variables are written as
\begin{align}
    \phi_i(t)\approx \hat{\phi}_i^\mu(t) &= \vartheta_\mu(t)(\omega_i-\Omega_\mu)+f_\mu(t), \label{eq:PhiApprox}\\
    \kappa_{ij}(t)\approx\hat{\kappa}_{ij}^{\mu\nu}(t)&=\varkappa_{\mu\nu}(t).\label{eq:KappaApprox}
\end{align}
In this ansatz each phase oscillator is parametrized by $\vartheta_\mu(t)$ describing the spread within the $\mu$th cluster with relative frequencies $(\omega_i-\Omega_\mu)$, where $\Omega_\mu=N_\mu^{-1} \sum_{i\in \mathcal{C}_\mu}\omega_i$ is the mean natural frequency of cluster $\mu$ with index set $\mathcal{C}_\mu$, and the collective phase $f_\mu(t)$ of each respective cluster. The coupling weights $\kappa_{ij}$ are assumed to be constant within each cluster and only vary across clusters. Our ansatz~\eqref{EQ:Phidot}--\eqref{EQ:Kappadot} changes the microscopic description $(\phi_i,\kappa_{ij})$ to a mesoscopic description for the clusters with the new collective coordinates $\vartheta_\mu$, $f_\mu$, and $\varkappa_{\mu\nu}$ \cite{Note3}. This reduces the high-dimensional system~\eqref{EQ:Phidot}--\eqref{EQ:Kappadot} from $N+N^2$ to $2M-1+M^2$ dimensions.

The equations of motion for the collective coordinates are obtained by minimizing the error made by the assumption~\eqref{eq:PhiApprox}--\eqref{eq:KappaApprox}, see~\cite{GOT15,SMI20,suppl} for details. For simplicity, we restrict ourselves in the following to the description of two clusters in the continuum limit $N\to\infty$ with fixed size ratios $n_1$ and $n_2=1-n_1$~\cite{Note4}. We introduce the order parameter for each cluster $r_{\mu}=\frac{1}{N} \left|\sum_{j\in \mathcal{C}_\mu} e^{ i\phi^\mu_j\left(t\right)}\right|$ that in the continuum limit is approximated by $r_{\mu}=\sin z/z$ with $z=\vartheta_\mu n_\mu/4$~\cite{KUR84,GOT15}. The resulting mesoscopic dynamics of system~\eqref{EQ:Phidot}--\eqref{EQ:Kappadot} is governed by
\begin{subequations}
\label{eq:CollCoords}
\begin{align}
	\begin{split}
		\dot{\vartheta}_\mu=&~1+\frac{48 \sigma}{n^2_\mu\vartheta_\mu}\left(\cos\left(\frac{\vartheta_\mu n_\mu}{4}\right)-r_\mu\right)\\
		&\times \left[\varkappa_{\mu\mu} n_\mu r_\mu+\varkappa_{\mu\nu}n_{\nu} r_{\nu}\cos f\right],
	\end{split}\label{eq:CollCoords1}\\
	\dot{f}=&~-\frac14-\sigma r_1 r_2 \left(n_1\varkappa_{21}+n_2\varkappa_{12}\right)\sin f,\\
	\dot{\varkappa}_{\mu\nu}=&~-\epsilon\left[\varkappa_{\mu\nu}+r_\mu r_\nu\sin\left(f_\mu-f_\nu+\beta\right)\right],\label{eq:CollCoords4}
\end{align}
\end{subequations}
with $\mu,\nu\in \{1,2\}$, $f=f_1-f_2$ is the phase difference of the two clusters. 
See~\cite{suppl} for details.  

\begin{figure}[t]
    \centering
    \includegraphics{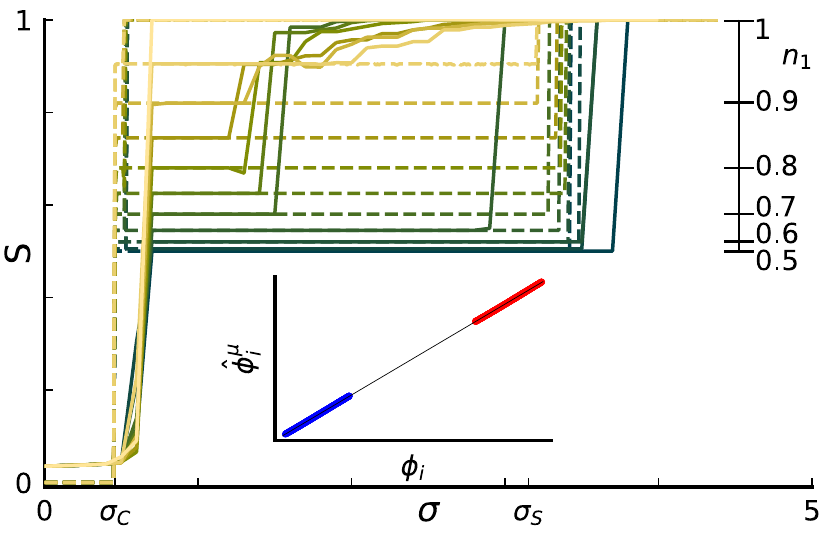}
    \caption{Bifurcation diagram for the adaptive Kuramoto model~\eqref{EQ:Phidot}--\eqref{EQ:Kappadot} (full system) with $N=1000$ oscillators (solid lines) and the reduced system \eqref{eq:CollCoords} (dashed lines), showing the synchronization index $S$ as a function of the coupling strength $\sigma$. For the full system a two-cluster state is initially prepared with a fraction of $n_1$ oscillators placed in the first cluster, and $n_2=1-n_1$ oscillators in the second cluster at $\sigma=1$, for the reduced system at $\sigma=0.5$ with values of $n_1$ from $0.5$ to $0.95$ in steps of $\Delta n_1=0.05$. Additionally the full system is simulated for $n_1=1$ to visualize hysteresis. 
    The analytical estimates of upper and lower bounds of two-cluster existence at $\sigma_S=3.152$ and $\sigma_C=0.460$ are marked on the axis. Inset: oscillator phases $\phi_i$ for the full system vs. the approximation $\hat{\phi}^\mu_i=\vartheta_\mu \omega_i$ of the reduced system for $n_1=0.5$ and $\sigma=2$; colors denote the two clusters. 
    Other parameters: $\hat{\omega}=0.25$, $\beta=-0.53\pi$ and $\epsilon=0.01$.}
    \label{fig:Figure3}
\end{figure}

Figure~\ref{fig:Figure3} shows a comparison of the high-dimensional adaptive Kuramoto system~\eqref{EQ:Phidot}--\eqref{EQ:Kappadot} and the reduced system~\eqref{eq:CollCoords}.
We use $N=1000$ oscillators with a fixed realization of natural frequencies to probe the continuum limit ($N\to\infty$) mean-field approach. For two-cluster configurations with varying relative number of oscillators in the first cluster from $n_1=0.5$ to $n_1=0.95$, we prepare special initial conditions that result in the desired state, see~\cite{suppl}. 
For the reduced system, we proceed analogously.


Figure~\ref{fig:Figure3} shows that the dynamics of the two-cluster state is captured fully by the collective coordinate framework. The solid lines (full system) overlap with the dashed lines (reduced system). The single-step transition, also seen in Fig.~\ref{fig:Figure1}(c), is well explained by the merging of two clusters in the reduced system. In both systems the multicluster structure ceases to exist beyond a certain coupling strength $\sigma_{c}$. The critical values for the onset of cluster $\sigma_c$ and full synchronization $\sigma_s$ are well approximated by a perturbative approach for $\epsilon\ll1$ to the reduced system \eqref{eq:CollCoords}. We obtain $\sigma_c\approx0.460$ and $\sigma_s\approx 3.152$, see~\cite{suppl}. In particular, the analytic result shows that multicluster states exist only for an intermediate range of $\sigma$, which agrees with the observations in Fig.~\ref{fig:Figure1}.

 
In the inset of Fig.~\ref{fig:Figure3} the excellent agreement between the oscillator phases of the full system and the collective coordinate ansatz~\eqref{eq:PhiApprox} is shown for $n_1=0.5$.


Fig.~\ref{fig:Figure3} shows that two-cluster states in the reduced system are stable for a much larger interval in $\sigma$ than in the full system. This discrepancy may be linked to the observation that the stability of multicluster states is mainly determined by intracluster links~\cite{FEK21}. 
Such intracluster effects are not captured by our mean-field ansatz. However, the reduced system provides important insights into the existence of partially synchronized clusters from which the stability can be studied numerically employing the full system. This is relevant to the two transition scenarios observed in Fig.~\ref{fig:Figure1}: for the reduced two-cluster system the values for which full synchronization emerges are larger the more equal the respective cluster sizes are with a maximum at $n_1=n_2=0.5$.
Hence a homogeneous collection of clusters of similar size will remain stable for a wide range of coupling strengths whereas heterogeneous nucleation with a dominant initial cluster will entrain further oscillators upon increasing the coupling strength.   


In summary, we have shown two qualitatively different transitions to synchronization induced by the interplay of an adaptive network structure and finite size inhomogeneities in the natural frequency distributions: single-step and multi-step transitions. The transition through multiclusters makes these two phenomena different to explosive synchronization reported for systems with single or multiple adaptive coupling weights~\cite{SKA13a,ZHA15a,AVA18}. In the multi-step transition, a single large cluster (nucleus) forms around an inhomogeneity in the frequency distribution and successively grows until full synchronization is reached. In contrast, in the single-step synchronization transition, multiple equally sized clusters (nuclei) form around multiple inhomogeneities, grow and coexist stably. Each cluster moves with its mean frequency, which results in a higher difference of the average phase velocity between the clusters than between two freely moving oscillators. This higher difference inhibits the synchronization of the clusters for a significant range in the coupling strengths. Hence, this explains the observed abrupt first-order transition to full synchronization for high coupling strengths.


The described nucleation phenomena are very similar to heterogeneous nucleation induced by local impurities known, e.g., from cloud formation~\cite{PRU10}, crystal growth~\cite{MUL01} or Ostwald ripening in equilibrium and nonequilibrium systems~\cite{SCH91}. Due to this relation, our results provide an intriguing bridge between synchronization transitions in finite-size dynamical complex networks and thermodynamic phase transitions where the finite-size induced inhomogeneities in the natural frequencies take the role of impurities. Our numerical investigation has been complemented by a mean-field theory capable of describing multicluster states in the presence of an arbitrary frequency distribution and adaptive coupling weights.
By this, we contribute to the research on mean-field models of coupled phase oscillators~\cite{OTT08} where only recently first steps have been undertaken to include adaptive coupling~\cite{GKO22}. Remarkably, our reduced mean-field model provides an excellent approximation of the macroscopic multicluster dynamics as well as the microscopic phase relations. The multi-step transition with a continually changing size of the main cluster (nucleus) and the importance of the stability of each cluster is only partially captured by the mean-field approach introduced in this work. This limitation, however, could be overcome by generalizing methods on partial synchronization in generalized Kuramoto systems~\cite{TAN97,JI13,OLM14a,SMI20}.

\clearpage
\onecolumngrid
\begin{center}
	\textbf{\large Supplemental Material on:\\ Heterogeneous nucleation in finite size adaptive dynamical networks}\\[.2cm]
	Jan Fialkowski$^{1}$, Serhiy Yanchuk$^{2,3}$, Igor M. Sokolov$^4$, Eckehard Sch\"oll$^{1,2,5}$, Georg A. Gottwald$^6$, and Rico Berner$^{1,4}$\\[.1cm]
	{\itshape $^{1}$Institute for Theoretical Physics, Technische Universit\"at Berlin, Hardenbergstr.\,36, 10623 Berlin, Germany\\
		$^{2}$Potsdam Institute for Climate Impact Research, Telegrafenberg A 31, 14473 Potsdam, Germany\\
		$^{3}$Institute of Mathematics, Humboldt University Berlin, 12489 Berlin, Germany\\
		$^{4}$Department of Physics, Humboldt Universit\"at zu Berlin, Newtonstraße 15, 12489 Berlin, Germany\\
		$^{5}$Bernstein Center for Computational Neuroscience Berlin, Humboldt Universit\"at, 10115 Berlin, Germany\\
		$^{6}$School of Mathematics and Statistics, University of Sydney, Camperdown NSW 2006, Australia\\}
\end{center}
\twocolumngrid
\setcounter{equation}{0}
\setcounter{figure}{0}
\setcounter{table}{0}
\setcounter{page}{1}
\renewcommand{\theequation}{S\arabic{equation}}
\renewcommand{\thefigure}{S\arabic{figure}}
\renewcommand{\thetable}{S\arabic{table}}
\renewcommand{\bibnumfmt}[1]{[S#1]}
\renewcommand{\citenumfont}[1]{#1}
\setcounter{secnumdepth}{1}

\section{Numerical schemes}
\subsection{Integration schemes}
For the numerical integration of the full adaptive Kuramoto system with $N=50$ oscillators, the simulations are performed using a fifth order Runge-Kutta method and the time step is chosen at every step by estimating the error via comparison with a Runge-Kutta of fourth order and enforcing its smallness.

For the numerical integration of the full adaptive Kuramoto system with $N=1000$ oscillators, we use a procedure that combines a classical Runge-Kutta scheme for the oscillator phases with Euler's method for the coupling weights. We use a fixed $\Delta t=0.01$ as the step size. Let $u_i$ be the state vector of our system at time $t=i\Delta t$. In order to calculate $u_{i+1}=u_i+k$, with $k$ being the vector of change of our state, we split the calculation into an oscillator $k_\phi$ and a coupling $k_\kappa$ part. Since $\epsilon\ll 1$, the dynamics of the coupling is slower than the dynamics of the phases. For the slow dynamics of the coupling we employ a simple Euler scheme with
\begin{equation*}
    k_\kappa = \dot{k}_\kappa\left(u_i\right)\Delta t.
\end{equation*}
For the fast oscillator part $k_\phi$ we use a more accurate fourth-order Runge-Kutta scheme. Here $\dot{k_\phi}\left(u\right)$ corresponds to the equation of motion~(2) of the main text. Using an Euler step for the coupling significantly speeds up the computation and allows for the extensive sweeping protocols over many realisations. This is justified since the dynamics is slow and the coupling can be assumed to be constant during the fine-scale Runge-Kutta steps for the fast oscillator dynamics. 

For the numerical integration of the full Kuramoto system with $N=50$ oscillators, the integration is performed using a fifth order Runge-Kutta method and the time step is chosen at every step by estimating the error via comparison with a Runge-Kutta of fourth order and enforcing its smallness.

\subsection{Parameter sweeping protocols}\label{sec:sweeps}
For the parameter sweeping in Fig.~1 of the main text, we use different protocols depending on the type of analysis done. For most parts of our study, we consider systems of $N=50$ adaptively coupled oscillators. Large systems with $N=1000$ are used to show the accuracy of our mean-field approach for two-cluster states, however, the sweeping approach has been adjusted as described in the subsequent section.

\emph{Up-sweep protocol:} For systems with $N=50$ oscillators with given natural frequencies, we initiate each sweep with random initial conditions at $\sigma=0$. Then we simulate the system for a total of $10^4$ time units. The last $7000$ time units of which are used to calculate the synchronization index $S$, see definition below. 

\emph{Down-sweep protocol:} The down-sweep is performed in a similar manner, starting from the fully synchronized state at $\sigma=6$, decreasing $\sigma$ in steps of $\Delta\sigma=-0.01$ after every $10^4$ time units and using the previously calculated state as the new initial condition. We analyse the down-sweep until $\sigma=0$.

\emph{Synchronization index:} For the calculation of $S$, we evaluate if the mean phase velocities $\left\langle\dot{\phi}_i\right\rangle=1/T\int_{T_0}^{T_0+T}\dot{\phi}_i(t)\mathrm{d}t$ of the oscillators are the same allowing for a small difference of $\left\langle\dot{\phi}_i\right\rangle-\left\langle\dot{\phi}_j\right\rangle<10^{-3}$ for transient time $T_0$ and sufficiently large averaging time $T$. The total number of pairings for which this is true divided by the total number of oscillator pairs $N^2$ determines the synchronization index $S$. The final state $\phi_i$ reached after these $10^4$ time units is used as the initial condition for the simulation at a value of the increased coupling strength $\sigma = \Delta\sigma=0.01$. This  process repeats during the up-sweep until $\sigma=6$.

Alternatively $S$ can be determined by analyzing the coupling matrix with elements $\kappa_{ij}$. For two oscillators with different dynamical frequencies, i.e., $\left\langle\dot{\phi}_i\right\rangle\neq\left\langle\dot{\phi}_j\right\rangle$, the coupling between them is close to zero in leading order $\kappa_{ij}=0+\mathcal{O}(\epsilon)$, see Ref.~[40] of the main text. If, however, the dynamical frequency is the same, then the coupling weight between oscillators tends asymptotically to $\kappa_{ij}=-\sin(\phi_i-\phi_j+\beta)$. The phase difference between the oscillators is approximately constant and small for synchronized clusters. Hence, for $\beta=-0.53\pi$, the coupling weight between two synchronized oscillators is approximately $\kappa_{ij} \approx 1$. This allows for the calculation of the synchronization index $S$ using a threshold for the coupling weights. We choose a threshold of $0.5$ so that $s_{ij}=1$ for $\kappa_{ij}>0.5$ and $s_{ij}=0$ otherwise. Calculating the mean of the resulting matrix yields $S$. Both methods for deriving $S$ are equivalent and we make use of either one or the other for computational convenience.

\subsection{Parameter sweeping for two-cluster states}
The scans through $\sigma$ shown in Fig.~3 of the main text for systems with $N=1000$ oscillators are achieved as follows. To make sure the analysis is performed for the desired two-cluster state with a particular $n_1$, the system is initiated with oscillators belonging to a cluster being grouped together and their intracluster coupling set to the maximal value $\varkappa_{\mu\mu}=-\sin\beta$, whereas the intercluster coupling is set to the minimal value $\varkappa_{\mu\nu}=0$, i.e., the intercluster coupling is initially switched off. The reduced system uses the same initial values for the couplings as well as $\vartheta_\mu=0.1$ for each cluster, with $f_\mu=0$ or $0.3\pi$ respectively.

With the desired two-cluster state the simulation is started at an estimated value of $\sigma$ that is close to the border of desynchronization. We choose $\sigma_0=0.5$ for the reduced system and $\sigma_0=1$ for the full system. With this, the protocols described above for sweeping through the values of $\sigma$ are applied with a step size of $\Delta\sigma=0.01$ for the reduced system and $\Delta\sigma=0.1$ for the full system. Due to its reduced complexity and the ten times higher resolution in $\sigma$, the reduced system for the collective coordinates only requires a shorter transient time of $T_0=500$ time units before the order parameter is calculated from the simulation of further $T=1500$ time units. 
The full phase oscillator system, Eqs.(1)--(2) of the main text, for $N=1000$ oscillators requires more care in choosing the integration times. Let $\sigma_J$ be the current value of $\sigma$ of the simulation after $J$ steps. First the system is run for a short time of $T_{1}=100$ time units and the order parameter is calculated over subsequent $T=1000$ time units. After this we calculate the change in the order parameter compared to the previous value at $\sigma_{J-1}$, $\Delta R_2(J)=\left\langle R_2\right\rangle\left(\sigma_J\right)-\left\langle R_2\right\rangle\left(\sigma_{J-1}\right)$. This change is compared with the change in the previous step $\Delta R_2(J-1)$. If $\Delta R_2 (J) < 2\Delta R_2 (J-1)$ we assume the state to be stable, proceed as outlined above and simulate the system at a higher value of $\sigma_{J+1}=\sigma_J+\Delta\sigma$. If on the other hand $\Delta R_2 (J) \geq 2\Delta R_2 (J-1)$, the system is assumed to undergo a change in its state and is simulated for a further $T_{2}=10^4$ time units to allow for relaxation towards its new stationary state before the order parameter is estimated. To ensure that the transition to the new state has not been missed due to the smaller transient times the previous state for $\sigma_{J-1}$ is simulated again for $T_{2}$ time units. Should this repeated simulation reveal the state to be unstable already at $\sigma_{J-1}$, we use the newly simulated state as an initial condition to continue the sweep with $\sigma_J$.

\section{Equiprobably drawn frequencies}
\begin{figure}[h]
    \centering
    \includegraphics[width=0.475\textwidth]{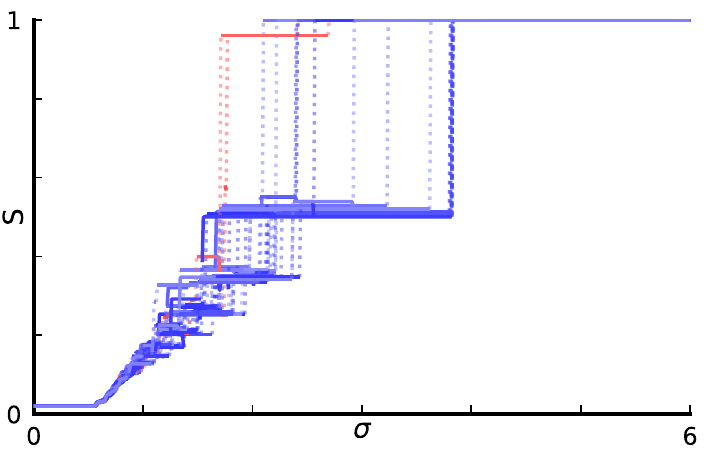}
    \caption{The synchronization index $S$ for increasing values of the coupling strength $\sigma$. The simulations use an equidistant distribution of natural frequencies $\omega_i=-0.25+0.5(i-1)/(N-1)$. Shown are $30$ simulations from different random initial conditions for $N=50$ oscillators each, where $\sigma$ is varied from $0$ to $6$ in steps of size $\Delta\sigma=0.01$. The protocol is the same as in Fig.~1(a) of the main text. Other parameters: $\beta=-0.53\pi$, $\epsilon=0.01$.}
    \label{fig:equidistant}
\end{figure}
To test if the occurrence of the two distinct scenarios is indeed caused by finite size inhomogeneities in the realization of the natural frequency distribution, we simulate the system with equiprobably drawn frequencies. In particular, we simulate (1)-(2) of the main text using an equidistant draw of the natural frequencies $\omega_i=-0.25+0.5(i-1)/(N-1)$ for $N=50$ oscillators. Figure~\ref{fig:equidistant} shows the results of $30$ simulations for different initial conditions. The system exhibits a high degree of multistability for different initial conditions leading to different cluster structures as $\sigma$ is increased. Most of the $30$ initial conditions follow the path of the single-step transition to synchrony, with two or three rather evenly sized clusters. For some initial conditions, however, we observe the formation of a single dominant cluster with mean-frequency close to zero leading to the multi-step transition. In the absence of finite size fluctuations in the realization of the natural frequencies, fluctuations in the initial conditions lead to the differentiation in the synchronization process into the different paths. Hence, when approximating the continuum limit without nucleation seeds stemming from inhomogeneities in the realization of the natural frequency distribution, typically homogeneous nucleation takes place with a single-step transition. 

\section{Multimodal frequency distributions}
\begin{figure}
    \centering
    \includegraphics[width=0.475\textwidth]{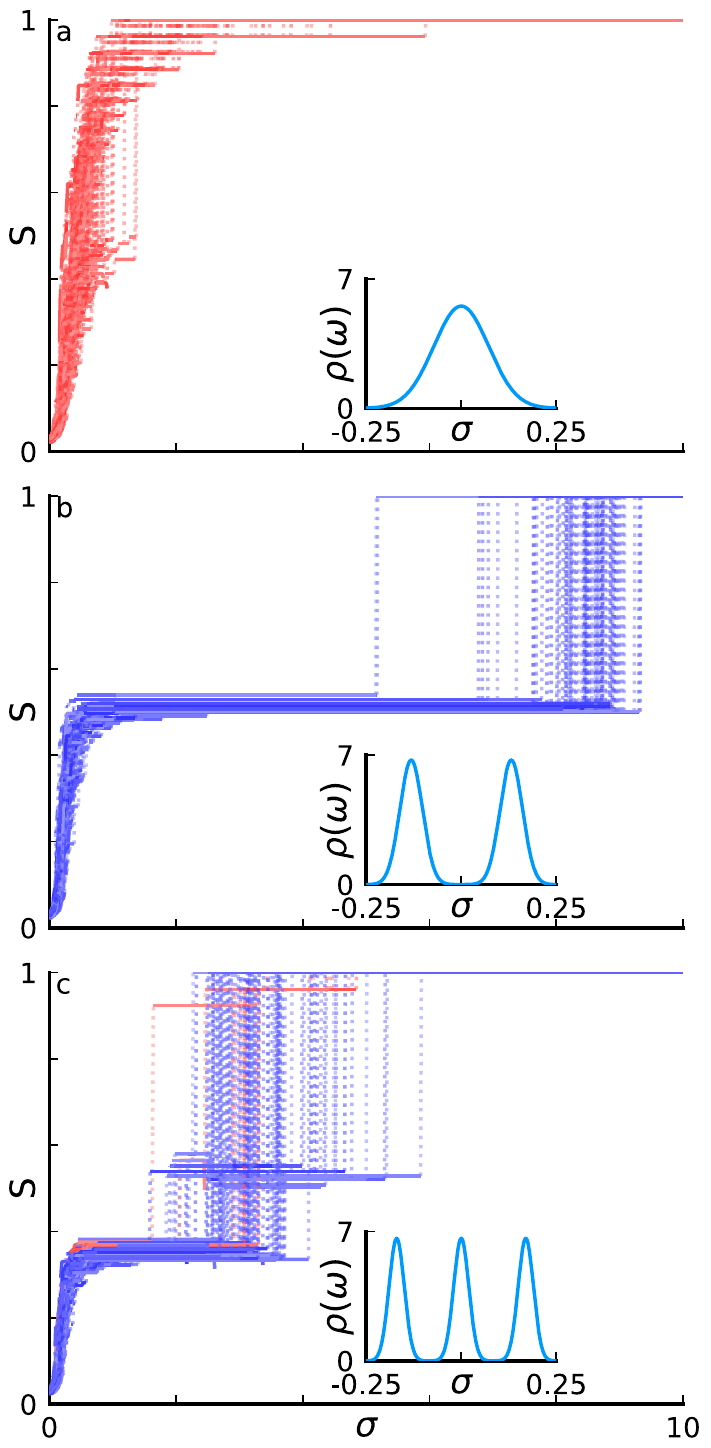}
    \caption{Paths to synchrony for different multimodal Gaussian distributions. The procedure for these plots is the same as in Fig.~1 of the main text (up-sweep). (a) Unimodal Gaussian distribution with a standard deviation of $s=0.0722$ and mean $\mu=0$, corresponding to the multi-step transition with a single synchronization seed. (b) Bimodal Gaussian distribution of equal standard deviation $s=0.0296$ with mean values of $\mu=\pm0.132$. This distribution corresponds to the case of the single-step transition with two synchronization seeds. (c) Trimodal Gaussian distribution of equal standard deviation $s=0.020$. One peak is centered at $0$ whereas the other two have a mean of $\mu=\pm0.170$, respectively. The insets show the respective underlying probability density function of the natural frequencies. Each panel consists of $100$ runs with different realizations of the natural frequencies and initial conditions.}
    \label{fig:Multimodal}
\end{figure}
In order to support the findings in the main text, we simulate the system~(1)--(2) of the main text for $N=50$ oscillators with different frequency distributions modelling the distributions displayed in the inset of Fig.~1(a) of the main text. To model different nucleation seeds we consider multimodal frequency distributions with each peak representing one nucleation site. In particular we consider a multimodal distribution consisting of $M$ separated Gaussians of the form
\begin{equation*}
    \rho\left(\omega\right)=\sum_{m=1}^{M}\frac{1}{M}\mathcal{N}\left(\mu_m,s\right).
\end{equation*}
Here $\mathcal{N}\left(\mu_m,s\right)$ is a normalized Gaussian distribution with mean $\mu_m$ and standard deviation $s$. 

The associated paths to synchronization are shown in Fig.~\ref{fig:Multimodal}. We examine distributions with up to $M=3$ and choose the parameters for the Gaussian distributions in such a way that the peaks are clearly separated and decay sufficiently fast within the interval $\omega\in\left[-0.25,0.25\right]$. The distributions are shown as insets in each panel. Fig.~\ref{fig:Multimodal}~(a) shows the resulting paths for a single peak with a standard deviation of $s=0.0722$. All realizations of the natural frequencies follow the path of the multi-step transition where a big cluster emerges around the mean of the frequency distribution that progressively grows by absorbing more of the outer oscillators. The case of two peaks with a standard deviation of $s=0.0296$ is shown in Fig.~\ref{fig:Multimodal}~(b). One can see clearly that all realisations follow the path of the single-step transition. As explained in the main text this is due to the emergence of frequency clusters of roughly equal size with mean cluster phase velocities symmetrical around $0$. These clusters merge abruptly at sufficiently large coupling strength $\sigma$ forming a single cluster. This corresponds to the observation that synchronization nuclei form around inhomogeneities in the realization of the natural frequency distribution and determine the transition path as $\sigma$ is increased. A single seed leads to the multi-step transition, whereas two seeds lead to the single-step transition. For a triple peak distribution, shown in fig.~\ref{fig:Multimodal}~(c), the synchronization process is more complex. It starts as expected with the creation of three frequency-synchronized clusters, but as the coupling strength is increased not all simulations show a sudden merging of all three clusters. In fact some of the simulations again show either the creation of a single dominant cluster leading to higher synchronization index $S$ or the emergence of two big clusters that subsequently merge at even higher coupling strengths $\sigma$.

\section{Hysteresis}
\begin{figure}[ht]
    \centering
    \includegraphics[width=\columnwidth]{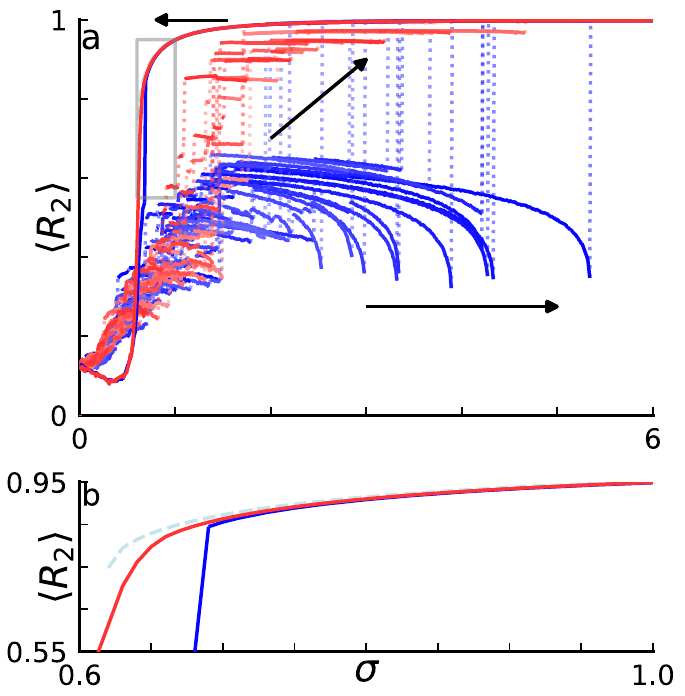}
    \caption{Hysteric transition in system \eqref{EQ:Phidot}--\eqref{EQ:Kappadot} with $N=50$ oscillators. (a) Up-sweep: Shown are the time averaged second moment of the order parameter $\langle R_2\rangle$ as a function of the coupling strength $\sigma$ for $100$ simulations with $N=50$ oscillators. Each run is initiated with random initial conditions and $\sigma$ is increased in steps of $\Delta\sigma=0.01$. For each run natural frequencies are drawn independently from a uniform distribution $\omega_i\in\left[-0.25,~0.25\right]$. Down-sweep: Additionally shown are the time averaged second moment of the order parameters for two simulations with decreasing $\sigma$ starting from the final state of the up-sweep protocol. The natural frequencies of these down-sweeps are chosen such that one corresponds to the single-step transition path and one to the multi-step transition path. The light-blue dashed line corresponds to the stable branch of the analytical solution of Eq.~\eqref{Bestimmung theta}. (b) We provide a blow-up for the region depicted as a grey rectangle in panel (a). Other parameters: $\beta=-0.53\pi$, $\epsilon=0.01$.}
    \label{fig:Hysteresis}
\end{figure}
In Figure~\ref{fig:Hysteresis} we show the same data as in Fig.~1 of the main text (up-sweep) as well as the behavior of the time averaged order parameter $\langle R_2\rangle$ as $\sigma$ is decreased via the down-sweep protocol. Additionally we show the analytical one-cluster solution derived from Eq.~\eqref{Bestimmung theta} as a dashed line. For a large range in $\sigma$, the synchronous states is shown to be co-stable with various multicluster states. The fully synchronized state is stable until small values of $\sigma$, where both down-sweep simulations show a sudden and sharp decline in their order parameter below a critical value of $\sigma$. As soon as the fully frequency synchronized state disappears most of the other coexisting multicluster states disappear as well, compare with Fig.~3 of the main text. The system abruptly approaches an asynchronous state. The hysteretic behavior presented in~\ref{fig:Hysteresis} is a common feature of first-order phase transitions. The analytical solution predicts the order parameter and the border to desynchronization very well, with a small error towards higher order parameters for lower values of $\sigma$. We attribute this error to the assumption of infinitely many oscillators used in the derivation of the analytic solution.

\section{Transitions for different values of $\beta$}
\begin{figure}[ht]
    \centering
    \includegraphics[width=\columnwidth]{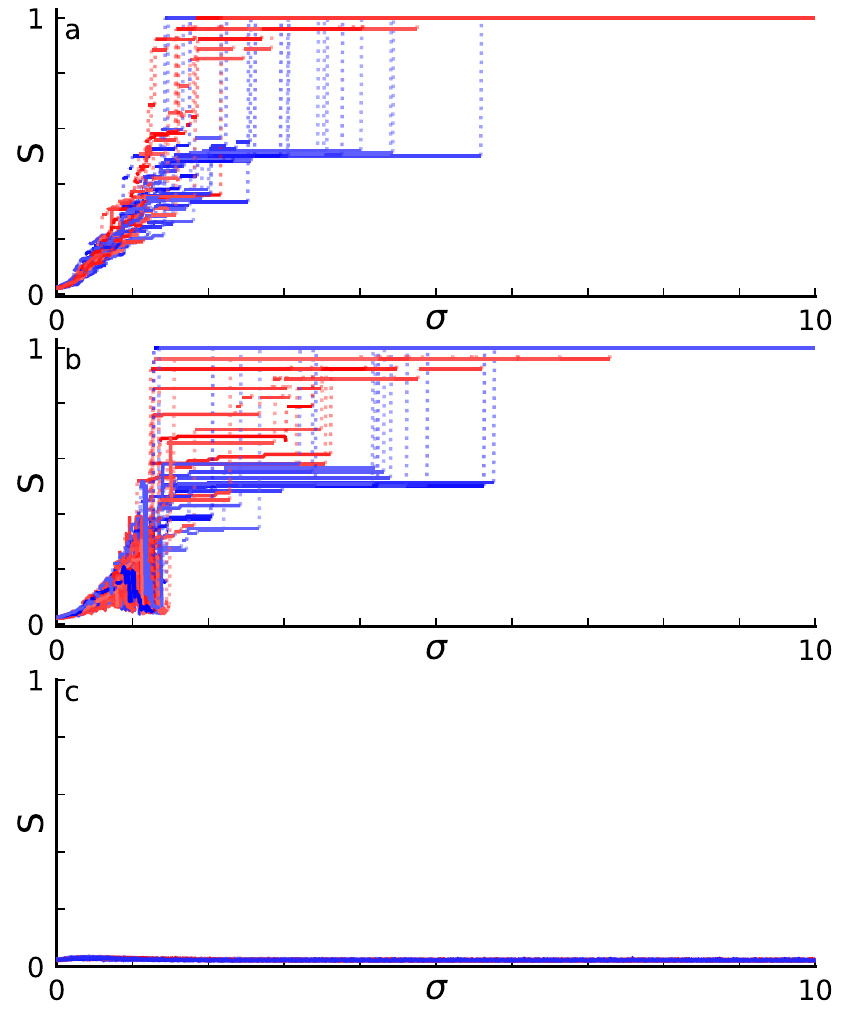}
    \caption{Paths to synchrony for system~\eqref{EQ:Phidot}--\eqref{EQ:Kappadot} for different values of $\beta$ (a) $\beta=-0.4\pi$, (b) $\beta=-0.25\pi$, (c) $\beta=0.25\pi$. Synchronization index $S$ as a function of the coupling strength $\sigma$ 
    for $30$ simulations with $N=50$ oscillators. Each run was initiated with random initial conditions and $\sigma$ was increased in steps of $\Delta\sigma=0.01$. For each run, natural frequencies are drawn independently from a uniform distribution $\omega_i\in\left[-\hat{\omega},~\hat{\omega}\right]$ with $\hat{\omega}=0.25$. The dotted lines indicate jumps in the synchronization index during the transition and colors represent the multi-step (red) and single-step (blue) transitions as in Fig.~1 of the main text. The synchronization index $S$ is determined with an averaging time window $T=10^4$ and transient time $T_0=4\cdot10^4$. Other parameters: $\epsilon=0.01$.}
    \label{fig:Betas}
\end{figure}
In order to examine the influence of $\beta$ we repeat the up-sweep originally presented in Fig.~1 of the main text for three different values of $\beta$. The results are presented in Fig.~\ref{fig:Betas}. The synchronization for $\beta=-0.4\pi$ is presented in Fig.~\ref{fig:Betas}~(a). The system still tends to full synchronization but requires higher values of the overall coupling strength $\sigma$ to reach the synchronized state. Moreover, the two transition paths reported for $\beta=-0.53\pi$ in the main text still exist for $\beta=-0.4\pi$. When $\beta$ is further increased to $\beta=-0.25\pi$ the two paths to synchronization are still visible, see Fig.~\ref{fig:Betas}~(b). We note that the details of the transitions might change, in particular the statistics for the natural frequencies $\omega_i$ can look different compared to the statistics presented in Fig.~1 of the main paper. Also, since different values for $\beta$ determine which phase configurations lead to high coupling strengths, the exact nature of the finite size effects might change. The examination of these effects is beyond the scope of this work and is an opportunity for future work. The results for the positive value $\beta=0.25\pi$ is presented in Figs.~\ref{fig:Betas}~(c). Here we do not observe any synchronization process which agrees with previous results, see [40] of the main text for details. We can thus conclude that the results presented in the main paper are observable for parameter values other than $\beta=-0.53\pi$.

\section{Antipodal symmetry}
In order to motivate why it is sufficient to plot the absolute values $|\kappa_{ij}|$ in Fig.~2 of the main text, we show that the system 
\begin{align}
	\frac{d\phi_i}{dt}&=\omega_i-\frac{\sigma}{N}\sum_{j=1}^N\kappa_{ij}\sin(\phi_i-\phi_j)\label{EQ:Phidot},\\
	\frac{d\kappa_{ij}}{dt}&=-\epsilon\left(\kappa_{ij}+\sin(\phi_i-\phi_j+\beta)\right)\label{EQ:Kappadot}
\end{align}
is invariant under the transformation 
\begin{align*}
		\phi_i&\rightarrow \tilde{\phi}_i=\phi_i+\pi,\\
    \kappa_{lm}&\rightarrow \tilde{\kappa}_{lm}=-\kappa_{lm}~\text{where either $l=i$ or $m=i$},
\end{align*}
for a fixed $i=1,\dots,N$.

For equation \eqref{EQ:Phidot} it is sufficient to examine the term in the sum which yields
\begin{align*}
    \tilde{\kappa}_{ij}\sin\left(\tilde{\phi}_i-\phi_j\right)&=-\kappa_{ij}\sin\left(\phi_i+\pi-\phi_j\right)\\
    &=\kappa_{ij}\sin\left(\phi_i-\phi_j\right).
\end{align*}
Similarly, equation \eqref{EQ:Kappadot} remains invariant under the transformation as can be readily seen by comparing with the transformed equation 
\begin{align*}
    -\frac{d\tilde{\kappa}_{ij}}{dt}=-\epsilon\left(-\tilde{\kappa}_{ij}-\sin(\tilde{\phi}_i-\phi_j+\beta)\right).
\end{align*}
This transformation can be applied to any of the $N$ oscillators, changing the sign of a particular $\kappa_{lm}$.

\section{Derivation of multicluster mean-field equations}
\label{S.VI}
As outlined in the main text, we use an ansatz that describes each of the $M$ clusters with three collective variables and reduces the oscillator-oscillator coupling to a cluster-cluster coupling. This is done by first assigning each oscillator $i$ to a cluster $\mu$ with $\mathcal{C}_\mu$ being the index set of all oscillators in cluster $\mu$. Second, we assume a set of collective coordinates $\vartheta_\mu(t), f_\mu(t)$ and $\varkappa_{\mu\nu}(t)$ that describe the temporal dependency of the phases and the coupling as follows
\begin{align}
    \phi_i(t)\approx \tilde{\phi}_i^\mu(t) &= \vartheta_\mu(t)(\omega_i-\Omega_\mu)+f_\mu(t), \label{eq:PhiApprox}\\
    \kappa_{ij}(t)\approx\tilde{\kappa}_{ij}^{\mu\nu}(t)&=\varkappa_{\mu\nu}(t).\label{eq:KappaApprox}
\end{align}
The dynamical description of each oscillator is split up into two parts. The variable $f_\mu$ serves as a collective phase variable for the cluster of the oscillator. The spread within the $\mu$th cluster with relative frequency $(\omega_i-\Omega_\mu)$ is described by $\vartheta_\mu$. The coupling variables $\kappa_{ij}$ are reduced from an oscillator-oscillator coupling to a cluster-cluster coupling $\varkappa_{\mu\nu}$. Note that Eqs.~\eqref{eq:PhiApprox}--\eqref{eq:KappaApprox} are equivalent to Eqs.~(3)--(4) of the main text.
Inserting the ansatz Eq.~\eqref{eq:PhiApprox}--\eqref{eq:KappaApprox} into the original system Eq.~\eqref{EQ:Phidot}--\eqref{EQ:Kappadot} yields a non-zero $N+N^2$ dimensional error vector $\mathbf{E}=\left(E_{\phi_{1}^1}\ldots E_{\phi_N^M}, E_{\kappa_{1,1}^{1,1}}\ldots E_{\kappa_{N,N}^{M,M}}\right)$.
Introducing the difference in natural frequencies $\Delta\omega_{ij}:=\omega_i-\omega_j$ and the phase difference $\Delta\phi_{ij}^{\mu\nu}:=\tilde{\phi}_i^\mu-\tilde{\phi}_j^\nu=\vartheta_\mu(\omega_i-\Omega_\mu)+f_\mu-\vartheta_\nu(\omega_j-\Omega_\nu)-f_\nu$ we write the error corresponding to the dynamics of the phase and of the coupling as
\begin{align*}
	\begin{split}
		E_{\phi_i^\mu}=&~\dot{\vartheta}_\mu\left(\omega_i-\Omega_\mu\right)+\dot{f}_\mu-\omega_i\\
		&+\frac{\sigma}{N}\sum_{j\in \mathcal{C}_\mu}\varkappa_{\mu\mu}\sin\left(\vartheta_\mu\Delta\omega_{ij}\right)\\
		&+\frac{\sigma}{N}\sum_{\nu\neq \mu}\sum_{j\in \mathcal{C}_\nu}\varkappa_{\mu\nu}\sin\left(\Delta\phi_{ij}^{\mu\nu}\right),
	\end{split}\\
	E_{\kappa_{ij}^{\mu\nu}}=&~\dot{\varkappa}_{\mu\nu}+\epsilon\left[\varkappa_{\mu\nu}+\sin\left(\Delta\phi_{ij}^{\mu\nu}+\beta\right)\right].
\end{align*}
The error is minimized by requiring that it is orthogonal to the manifold of the ansatz \eqref{eq:PhiApprox}--\eqref{eq:KappaApprox} parametrized by the collective coordinates. The tangent space is spanned by the partial derivatives of the ansatz functions with respect to the collective coordinates. 
In particular, if we denote the ansatz~\eqref{eq:PhiApprox}--\eqref{eq:KappaApprox} as $u=\left(\tilde{\phi}, \tilde{\kappa}\right)$ and the collective coordinates $\mathbf{c}=\left(\vartheta_\mu, f_\mu, \kappa_{\mu\nu}\right)$, the minimization of the error amounts to  $\left(\frac{\partial}{\partial c_i}u\right)^T\cdot\mathbf{E}=0$ for all $i=1,2,3$. This yields the desired temporal evolution equations for the collective coordinates. 
Projecting onto $\partial u/\partial {\vartheta_\mu}$ yields
\begin{multline}
		\dot{\vartheta}_\mu\sum_{i\in \mathcal{C}_\mu}\left(\omega_i-\Omega_\mu\right)^2 =~\sum_{i\in \mathcal{C}_\mu}\left(\omega_i-\Omega_\mu\right)\omega_i\\
		-\frac{\sigma}{N}\sum_{i\in \mathcal{C}_\mu}\left(\omega_i-\Omega_\mu\right)\sum_{j\in \mathcal{C}_\mu}\varkappa_{\mu\mu}\sin\left(\vartheta_\mu\Delta\omega_{ij}\right)\\
		-\frac{\sigma}{N}\sum_{i\in \mathcal{C}_\mu}\left(\omega_i-\Omega_\mu\right)\sum_{\nu\neq \mu}\sum_{j\in \mathcal{C}_\nu}\varkappa_{\mu\nu}\sin\left(\Delta\phi_{ij}^{\mu\nu}\right).
	\label{eqm_vartheta_full}
\end{multline}
Here, we used that, by definition, $\sum_{i\in\mathcal{C}_\mu}\left(\omega_i-\Omega_\mu\right)=0$ to eliminate $\dot{f}_\mu$ from this equation. Similarly, the projections onto $\partial u/\partial {f_\mu}$ and onto $\partial u/\partial {\kappa_{\mu\nu}}$ yield  
\begin{align}
	\begin{split}
		\dot{f}_\mu &=~\Omega_\mu-\frac{\sigma}{N_\mu N}\sum_{i,j\in \mathcal{C}_\mu}\varkappa_{\mu\mu}\sin\left(\vartheta_\mu\Delta\omega_{ij}\right)\\
		&-\frac{\sigma}{N_\mu N}\sum_{i\in \mathcal{C}_\mu}\sum_{\nu\neq \mu}\sum_{j\in \mathcal{C}_\nu}\varkappa_{\mu\nu}\sin\left(\Delta\phi_{ij}^{\mu\nu}\right),
	\end{split}\\
	\dot{\varkappa}_{\mu\nu}&=-\epsilon\left[\varkappa_{\mu\nu}+\frac{1}{N_\mu N_\nu}\sum_{i\in \mathcal{C}_\mu}\sum_{j\in \mathcal{C}_\nu}\sin\left(\Delta\phi_{ij}^{\mu\nu}+\beta\right)\right].
		\label{eqm_varkappa_full}
\end{align}
Equations~\eqref{eqm_vartheta_full}-\eqref{eqm_varkappa_full} approximate the dynamics for a finite number of oscillators and can be analyzed for any given set of natural frequencies. To further simplify these equations, we consider the continuum limit $N\to\infty$, where sums of the form $\frac{1}{N}\sum_{i=1}^N g(\omega_i)$ are evaluated as $\int g(\omega) \rho\left(\omega\right) d\omega$ for an arbitrary function $g$. We mainly consider a uniform frequency distribution with
\begin{equation*}
    \rho\left(\omega\right)=\begin{cases}2,\text{ for }\omega\in\left[-0.25,0.25\right],\\
        0,\text{ otherwise.}\end{cases}
\end{equation*}

For the analysis of two cluster states we consider a distribution with disjoint support. We consider  
\begin{align*}
	\rho_1(\omega)&=\begin{cases}\frac{2}{n_1} \text{  for } \omega\in\left[-0.25,-0.25+0.5n_1\right],\\
        0\text{ otherwise,}\end{cases}\\
	\rho_2(\omega)&=\begin{cases}\frac{2}{n_2}\text{  for } \omega\in\left[0.25-0.5n_2,0.25\right].\\
        0\text{ otherwise,}\end{cases}
\end{align*}
with $n_2=1-n_1$. The $N_1=n_1 N$ oscillators with frequencies drawn from $\rho_1(\omega)$ are assigned to cluster $1$, and the remaining $N_2=n_2 N$ oscillators with frequencies drawn from $\rho_2(\omega)$ are assigned to cluster $2$. The probability density of the full system can be recovered as a mixture of the two cluster probability densities $\rho\left(\omega\right)=n_1\rho_1\left(\omega\right)+n_2\rho_2\left(\omega\right)$ In this case the order parameter for a single cluster is $r_\mu=\frac{4}{n_\mu \vartheta_\mu}\sin\left(\frac{n_\mu \vartheta_\mu}{4}\right)$. Introducing $f=f_1-f_2$, we arrive at equations (5a)-(5c) in the main text for the collective coordinates which we recall here for completeness 
\begin{subequations}
\label{eq:CollCoords}
\begin{align}
	\begin{split}
		\dot{\vartheta}_\mu=&~1+\frac{\sigma}{\mathrm{v}_\mu\vartheta_\mu}\left(\cos\left(\frac{\vartheta_\mu n_\mu}{4}\right)-r_\mu\right)\\
		&\times \left[\varkappa_{\mu\mu} n_\mu r_\mu+\varkappa_{\mu\nu}n_{\nu} r_{\nu}\cos f\right],
	\end{split}\\
	\dot{f}=&~\Omega_{12}-\sigma r_1 r_2 \left(n_1\varkappa_{21}+n_2\varkappa_{12}\right)\sin f,\\
	\dot{\varkappa}_{\mu\nu}=&~-\epsilon\left[\varkappa_{\mu\nu}+r_\mu r_\nu\sin\left(f_\mu-f_\nu+\beta\right)\right],
\end{align}
\end{subequations}
with $\Omega_{12}=\Omega_1-\Omega_2$ and the variance of the natural frequencies of a cluster $\mathrm{v}_\mu=\frac{1}{N_\mu}\sum_{i\in\mathcal{C}_\mu}\left(\omega_i-\Omega_\mu\right)^2$. Considering the continuum limits for the mean frequencies, i.e., $\Omega_\mu=\int\rho_\mu(\omega) \omega \,\mathrm{d}\omega$ and the frequency variance within a cluster, i.e., $v_\mu = \int\rho_\mu(\omega) (\omega-\Omega_\mu)^2 \,\mathrm{d}\omega$, we find $\Omega_1=(n_1-1)/4$, $\Omega_2=n_1/4$ and $v_\mu=n_\mu^2/48$. This yields, together with Eqs.~\eqref{eq:CollCoords}, the final mean-field Eqs. (5) of the main text.

\section{Perturbative approximation for the two-cluster state}
In this section, we derive an explicit expression for the two-cluster states emerging in~\eqref{eq:CollCoords}. To do this, we use a perturbative approach with respect to the small parameter $\epsilon$. Numerical simulations show that intercluster interactions, mediated via the coupling $\varkappa_{\mu\nu}$, can be considered as small. We assume that each cluster in first approximation can be treated as if it were isolated up to a small perturbation due to intercluster interaction. Further, we assume that the phase difference between the two clusters $f$ grows linearly in time corresponding to a relative rotational motion with constant phase velocity $\Omega^\prime$. This suggests the following expansion of the collective coordinates in orders of $\epsilon$
\begin{align}
\label{eq:pertAnsatz}
	\vartheta_\mu\left(t\right)&=\vartheta_\mu^{(0)}+\epsilon\vartheta_\mu^{(1)}\left(t\right)+\mathcal{O}\left(\epsilon^2\right),\\
	\varkappa_{\mu\nu}\left(t\right)&=\varkappa_{\mu\nu}^{(0)}\left(t\right)+\epsilon\varkappa_{\mu\nu}^{(1)}\left(t\right)+\mathcal{O}\left(\epsilon^2\right),\\
	f\left(t\right)&=\Omega^\prime t+\epsilon f^{(1)}\left(t\right)+\mathcal{O}\left(\epsilon^2\right).
\end{align}
Using a Taylor expansion, the order parameter is expanded as
\begin{align*}
	r_\mu\left(t\right)&=\frac{4}{\vartheta_\mu^{(0)} n_\mu} \sin\left(\frac{\vartheta_\mu^{(0)} n_\mu}{4}\right)\\
	&+\epsilon\frac{\vartheta_\mu^{(1)}}{\vartheta_\mu^{(0)}}
	\left[ \cos\left(\frac{\vartheta_\mu^{(0)} n_\mu}{4}\right)-\frac{4}{\vartheta_\mu^{(0)} n_\mu}\sin\left(\frac{\vartheta_\mu^{(0)} n_\mu}{4}\right)\right]\\
	&+\mathcal{O}\left(\epsilon^2\right)\\
	&=r_\mu^{(0)}+\epsilon r_\mu^{(1)}\left(t\right)+\mathcal{O}\left(\epsilon^2\right).
\end{align*}
In order to derive equations for the expansion coefficients, we insert the expansion~\eqref{eq:pertAnsatz} into Eqs.\eqref{eq:CollCoords} and equate powers of $\epsilon$. For the coupling variables this yields
\begin{align*}
	\dot{\varkappa}_{\mu\nu}^{(0)}&=0,\\	
	\dot{\varkappa}_{\mu\nu}^{(1)}&=-\left(\varkappa_{\mu\nu}^{(0)}+r_\mu^{(0)} r_\nu^{(0)}\sin\left(\pm \Omega^\prime t+\beta\right)\right),
\end{align*}
where $\Omega^\prime$ and $-\Omega^\prime$ correspond to $\varkappa_{12}$ and $\varkappa_{21}$, respectively. Note that for $\nu=\mu$ we set $\Omega^\prime =0$. To ensure constant intracluster coupling we require $\varkappa_{\mu\mu}^{(0)}=-\left(r_\mu^{(0)}\right)^2\sin\left(\beta\right)$. Setting $\kappa_{\mu\nu}^{(0)}=0$ ensures that the intercluster coupling is small compared to the intracluster coupling. For $\mu\ne\nu$ the intercluster coupling in the non-synchronized case is readily integrated yielding
\begin{align}
	\varkappa_{\mu\nu}^{(1)}(t) &= \frac{\pm 1}{\Omega^\prime}r_\mu^{(0)} r_\nu^{(0)}\cos\left(\pm\Omega^\prime t+\beta\right).
\end{align}  
Similarly, we find for $\vartheta_\mu$
\begin{align}
	\begin{split}
		0 &= 1+\frac{\sigma}{\mathrm{v}_\mu \vartheta_\mu^{(0)}}\left(\cos\left(\frac{\vartheta_\mu^{(0)} n_\mu}{4}\right)-r_\mu^{(0)}\right)\\
		&\cdot\left[\varkappa_{\mu\mu}^{(0)} n_\mu r_\mu^{(0)}+\varkappa_{\mu\nu}^{(0)} n_\nu r_\nu^{(0)}\cos\left(\Omega^\prime t\right)\right],
	\end{split}\label{lowerboundscheck}\\
	\begin{split}
    	\dot{\vartheta}_\mu^{(1)}&= \frac{\vartheta_\mu^{(1)}\sigma}{\mathrm{v}_\mu\left(\vartheta_\mu^{(0)}\right)^2}\left(\frac{8\sin\left(\frac{\vartheta_\mu^{(0)}n_\mu}{4}\right)}{\vartheta_\mu^{(0)} n_\mu}-2\cos\left(\frac{\vartheta_\mu^{(0)}n_\mu}{4}\right)\right.\\
    	&\left.-\frac{\vartheta_\mu^{(0)}n_\mu\sin\left(\frac{\vartheta_\mu^{(0)}n_\mu}{4}\right)}{4}\right)\\
    	&\cdot\left[n_\mu\kappa_{\mu\mu}^{(0)}r_\mu^{(0)}+\varkappa_{\mu\nu}^{(0)} n_\nu r_\nu^{(0)}\cos\left(\Omega^\prime t\right)\right]\\
    	&+\frac{\sigma}{\mathrm{v}_\mu \vartheta_\mu^{(0)}}\left(\cos\left(\frac{\vartheta_\mu^{(0)} n_\mu}{4}\right)-r_\mu^{(0)}\right)\\
    	&\cdot\left[\varkappa_{\mu\mu}^{(1)} n_\mu r_\mu^{(0)}+\varkappa_{\mu\mu}^{(0)} n_\mu r_\mu^{(1)}+\varkappa_{\mu\nu}^{(1)} n_\nu r_\nu^{(1)}\cos\left(\Omega^\prime t\right)\right].
	\end{split}\label{eq:Shapeperturbation}
\end{align}
A time-independent $\vartheta_\mu^{(0)}$ can only be achieved by setting $\varkappa_{\mu\nu}^{(0)}=0$ as above. Using these solutions to calculate the intercluster dynamics given by $f$ yields
\begin{equation}
	\begin{split}
		\epsilon \dot{f}^{(1)}&=\Omega_1-\Omega_2-\Omega^\prime\\
		&-\frac{\epsilon\sigma}{\Omega^\prime}\left(r_1^{(0)}r_2^{(0)}\right)^2\left[\frac{n_2}{2}\left(\sin\left(2\Omega^\prime t+\beta\right)-\sin\beta\right)\right.\\
		&\hspace{20pt}\left.+\frac{n_1}{2}\left(\sin\left(\beta-2\Omega^\prime t\right)-\sin\beta\right)\right].
	\end{split}
\end{equation}
For self-consistency with the ansatz~\eqref{eq:pertAnsatz}, we choose $\Omega^\prime$ in such a way that only sinusoidal terms are left in the previous equation. All dynamics which is linear in time is absorbed providing us with an explicit formula to derive $\Omega^\prime$, which reads
\begin{equation}
	0=\left(\Omega^\prime\right)^2+\Omega^\prime\left(\Omega_2-\Omega_1\right)-\frac{\epsilon\sigma}{2}\left(r_1^{(0)} r_2^{(0)}\right)^2\sin\beta.\label{Est.OmegaPrime}
\end{equation}

\section{Estimating upper and lower bounds for cluster synchronization}
Eq.~\eqref{Est.OmegaPrime} can be used to estimate the upper bound for the existence of the two-cluster solution. Solving the quadratic equation for real valued $\Omega^\prime$ implies that
\begin{equation}\label{eq:twoClexistsCond}
    \left(\Omega_1-\Omega_2\right)^2\geq -2\epsilon\sigma\left(r_1^{(0)} r_2^{(0)}\right)^2\sin\beta.
\end{equation}
For the chosen parameter values of $\epsilon=0.01$, $\beta=-0.53\pi$ and $\Omega_1-\Omega_2=-0.25$, we calculate the highest value of $\sigma$ for which the condition is no longer satisfied. A rough estimate using $r_1^{(0)}=r_2^{(0)}=1$ yields $\sigma_S=3.14$. Note that the approximate values for $r_\mu^{(0)}$ depend on $n_1$ and $\sigma$ explicitly, and implicitly through $\vartheta_\mu^{(0)}$, see Eq.~\eqref{lowerboundscheck}. For increasing values of $\sigma$ starting at our rough estimate, we find the value for $\sigma$ for which condition~\eqref{eq:twoClexistsCond} is no longer satisfied. This turns out to be about $\sigma_S\approx3.152$ for both $n_1=0.5$ and $n_1=0.95$ with an accuracy of $\Delta\sigma=0.001$.

To estimate a lower bound for the cluster synchronization we use Eq.~\eqref{lowerboundscheck} and analyse if it is solvable. Again checking for decreasing values of $\sigma$ using a step size of $\Delta\sigma=0.001$, we arrive at $\sigma_C\approx0.460$ as a lower bound for the two-cluster states. Here again the same value is attained for both $n_1=0.5\text{ and }n_1=0.95$.

Note that a similar analysis for the bounds is also feasible for a finite realization of the frequency distribution from Fig.~1 of the main text and the general reduced dynamical equations derived in Sec.~VII of the Supplemental Material. This would lead to realization-dependent estimates. The estimates would be quantitatively different but qualitatively the same.

\section{Single Cluster synchronization with collective coordinates}
In the case when all oscillators are frequency synchronized, we can use a simpler collective coordinate ansatz with just a single variable. The ansatz is as follows
\begin{align}
    \phi_i&\approx\hat{\phi}_i=\vartheta\omega_i\\
    \kappa_{ij}&\approx-\sin\left(\phi_i-\phi_j+\beta\right).
\end{align}
Here $\vartheta$ describes the distance of a given oscillator with natural frequency $\omega_i$ from the mean phase of the frequency synchronized cluster and $\kappa_{ij}$ are approximated by their steady-state values. 
Proceeding analogously as in Section~\ref{S.VI} we determine the error introduced by the ansatz of the phases and couplings, and obtain
\begin{align*}
	E_{\phi_i}&=\dot{\vartheta}\omega_i-\omega_i-\frac{\sigma}{N}\sum_{j=1}^{N}\sin\left(\vartheta\Delta\omega_{ij}+\beta\right)\sin\left(\vartheta\Delta\omega_{ij}+\alpha\right)\\
	E_{\kappa_{ij}}&=\dot{\vartheta}\Delta\omega_{ij}\cos\left(\vartheta\Delta\omega_{ij}+\beta\right).
\end{align*}
Calculating the projection of the error on the tangent space of the ansatz space and forcing it to be zero yields an equation of motion for our collective coordinate
\begin{multline}
	\dot{\vartheta}\left(\sum_{i=1}^{N}\omega_i^2-\sum_{i,j=1}^{N}\Delta\omega_{ij}^2\cos^2\left(\vartheta\Delta\omega_{ij}+\beta\right)\right)= \\
	\sum_{i=1}^{N}\omega_i^2+\frac{\sigma}{N}\sum_{i=1}^{N}\omega_i\sum_{j=1}^{N}\sin(\vartheta\Delta\omega_{ij}+\beta)\sin(\vartheta\Delta\omega_{ij}).
\end{multline}
Setting $\dot{\vartheta}=0$ yields a condition for the existence of a fully synchronized state. We obtain
\begin{equation}
	0=1-\frac{\sigma}{\mathrm{v}^2N^2}\sum_{i}\omega_i\sum_{j}\frac{1}{2}\cos\left(2\vartheta\Delta\omega_{ij}+\beta\right).\label{eqm_vartheta_singlecluster}
\end{equation}
Here $v=\frac{1}{N}\sum_{i=1}^N \omega_i$ is the standard deviation of the natural frequencies. Again choosing a uniform distribution of frequencies between $-0.25$ and $0.25$ provides a simplified equation to determine $\vartheta$
\begin{equation}
    1=48\sigma\sin(\beta)\frac{\sin\left(\frac{\vartheta}{2}\right)}{\vartheta}\frac{\left(\frac{\vartheta}{2}\right)\cos\left(\frac{\vartheta}{2}\right)-\sin\left(\frac{\vartheta}{2}\right)}{\vartheta^2}. \label{Bestimmung theta}
\end{equation}
This equation can be used to analyze the transition from a fully synchronized system to a desynchronized one. In Fig.~\ref{fig:Hysteresis}, we plot the stable branch of the analytical solution for the single cluster state along with the simulation result for the full system. Note that~\eqref{Bestimmung theta} for the reduced system gives rise to one stable and one unstable solution that vanish in a fold bifurcation, see Ref.~[58] from the main text. We observe a very good agreement of the numerical and the analytical results. Together with the findings presented in the two previous sections, the emergence of various transition pathways to synchrony and the hysteretic behavior can be accurately explained.

\section{Collective Coordinates compared with the full system}
\begin{figure}[ht]
    \centering
    \includegraphics{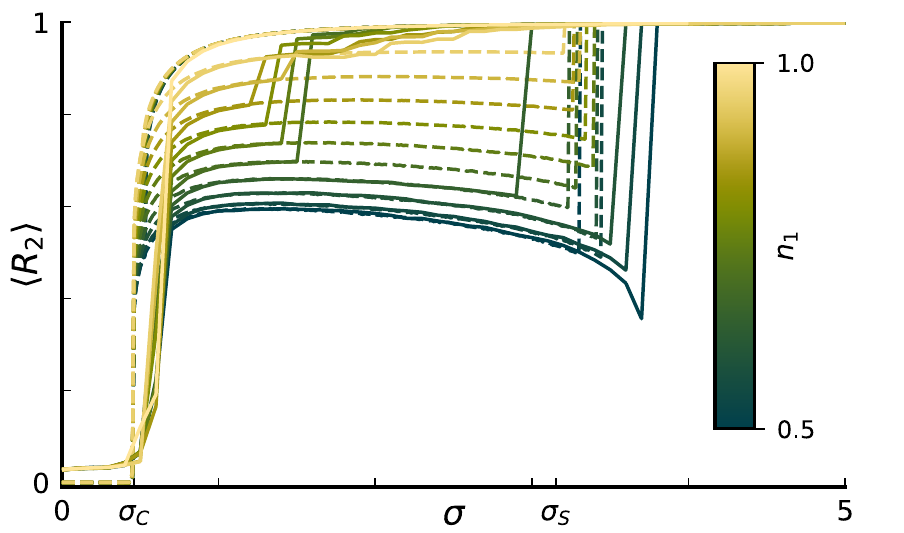}
    \caption{Bifurcation diagram for the adaptive Kuramoto model~\eqref{EQ:Phidot}--\eqref{EQ:Kappadot} (full system) with $N=1000$ oscillators (solid lines) and the reduced system \eqref{eq:CollCoords} (dashed lines), showing the time average (over $T=1000$) of the second Kuramoto-Daido order parameter $R_2$ 
    as a function of the coupling strength $\sigma$. For the full system a two-cluster state is initially prepared with a fraction of $n_1$ oscillators (denoted by color) placed in the first cluster, and $n_2=1-n_1$ oscillators in the second cluster at $\sigma=1$, for the reduced system at $\sigma=0.5$ with values of $n_1$ from $0.5$ to $0.95$ in steps of $\Delta n_1=0.05$. Additionally the full system is simulated for $n_1=1$ to visualize hysteresis. 
    The analytical estimates of upper and lower bounds of two-cluster existence at $\sigma_S=3.152$ and $\sigma_C=0.460$ are marked on the axis. 
    Other parameters: $\hat{\omega}=0.25$, $\beta=-0.53\pi$ and $\epsilon=0.01$.}
    \label{fig:Figure3}
\end{figure}
Figure~\ref{fig:Figure3} shows the second Kuramoto-Daido order parameter for simulations presented in Fig.~3 of the main paper. While the synchronization index $S$ and the inset in Fig.~3 show that the microscopic dynamics is well captured by the reduced system, the second order parameter in Fig.~\ref{fig:Figure3} additionally shows that the macroscopic inter-cluster dynamics are very well approximated. For low values of $\sigma$, the value of $\left\langle R_2\right\rangle$ grows due to the higher coherence of the individual clusters. Beyond a certain point, however, the value of the order parameter decreases. This is due to the increased interaction between the clusters that slows down their relative motion around a phase difference of $f=0.5\pi$. This leads to  lower values of $R_2$ at those time moments. For the case $n_1=0.5$, this particular configuration (phase difference $0.5\pi$) even yields $R_2=0$. As $\sigma$ is increased, the clusters spend more time with phase difference $f=0.5\pi$. This leads to the observed drop in the order parameter with increasing $\sigma$, which is particularly pronounced for cluster configurations near $n_1=0.5$. Since our approach reproduces the microscopic dynamics and the values for $R_2$ quite well, we can conclude that the cluster interaction is also well captured.

The reduced system provides important insights into the existence of partially synchronized clusters from which the stability can be studied numerically employing the full system. This is relevant to the two transition scenarios observed in Fig.~1 of the main text: the reduced two-cluster system withstands full synchronization for larger values of the coupling strength, the more equal the respective cluster sizes are, with a maximum at $n_1=n_2=0.5$. Hence a homogeneous collection of clusters of similar size will remain stable for a wide range of coupling strengths whereas heterogeneous nucleation with a dominant initial cluster will entrain further oscillators upon increasing the coupling strength. In Fig.~\ref{fig:Figure3} and Fig.~3 of the main text, it can be seen that structures with e.g. $n_1=0.7$ are destabilized much sooner than their almost evenly sized counterparts. In all cases that we have observed, intermediate cases are destabilized for smaller coupling constants and lead to a multi-step transition.

\section{Frequency clustering in the synchronization transition}
\textbf{\begin{figure}[ht]
    \centering
    \includegraphics[width=0.475\textwidth]{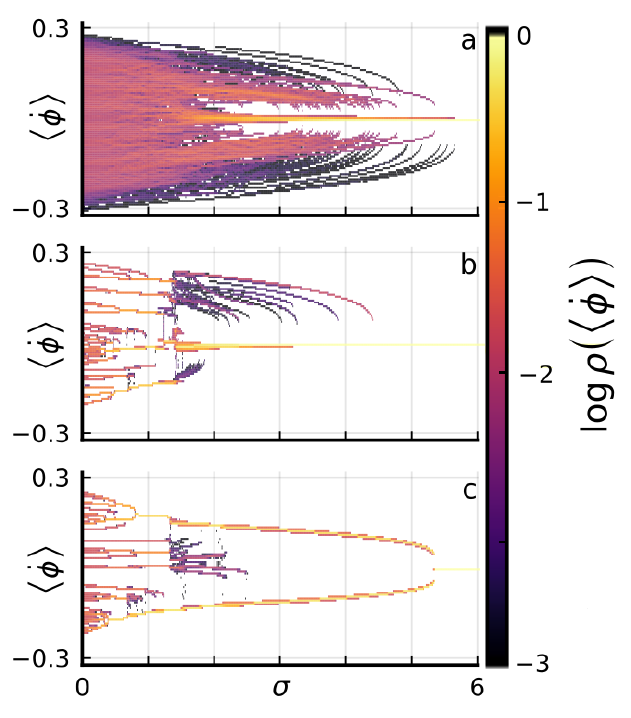}
    \caption{Evolution of the mean phase velocity $\left\langle\dot{\phi}_i\right\rangle$ as a function of increasing $\sigma$. For each value of $\sigma$ the color code denotes the natural logarithm of the relative number of oscillators found at a specific mean phase velocity. Each picture shows results for  $100$ simulations with $50$ oscillators each. (a) $100$ different realizations of natural frequencies, (b) using fixed frequencies with $100$ different initial conditions leading to a multi-step transition, (c) frequencies leading to the single-step transition. In the plot the frequencies are shifted vertically by the mean value of the realization of the natural frequencies such that each simulation is centered around $0$. The simulation procedure is the same as in Fig.~1 of the main text. Other parameters: $\beta=-0.53\pi$, $\epsilon=0.01$.}
    \label{fig:FigureVock}
\end{figure}}
To examine the clustering during the transition to synchrony in more detail, we present the distribution of the mean phase velocities $\rho\left(\left\langle\dot{\phi}_i\right\rangle\right)$ for $100$ different simulations encoded in color versus the overall coupling strength $\sigma$ in Fig.~\ref{fig:FigureVock}. The results are displayed in logarithmic scale. Each picture uses the data from $100$ simulation runs and for each run the mean phase velocity is shifted by the average frequency of that particular run. We use the up-sweep protocol described in Sec.~\ref{sec:sweeps} with a transient time of $T_{0}=10^5$ time units and an averaging time of $T=10^4$ time units. Figure~\ref{fig:FigureVock}(a) shows the statistics from $100$ simulations with different natural frequencies from a uniform distribution $\omega_i\in\left[-0.25,0.25\right]$ for each simulation. Figure~\ref{fig:FigureVock}(b) displays the data from $100$ simulations using a fixed realization of natural frequencies that leads to a multi-step first-order transition, similar to the ones shown in Fig.~1(b) of the main paper. The fixed realization leads to predetermined paths and allows for an analysis of the synchronization process in more detail. Note the emergence of a big group of oscillators around $\left\langle\dot{\phi}_i\right\rangle=0$ for small values of $\sigma$. This corresponds to the emergence of a large cluster for all simulations, exactly as shown in Fig.~2 of the main text. Figure~\ref{fig:FigureVock}(c) shows the data for $100$ simulations of a fixed realization of natural frequencies that lead to a single-step first-order transition, similar to the ones shown in Fig.~1(c) of the main paper. Note the two branches of oscillators located symmetrically around $0$. These represent the two clusters of almost equal size. Until these two clusters merge there are no oscillators with mean phase velocities around $0$. These two clusters are also observable in Fig.~2(g) of the main text.


\begin{thebibliography}{10}
\expandafter\ifx\csname url\endcsname\relax
  \def\url#1{{\tt #1}}\fi
\expandafter\ifx\csname urlprefix\endcsname\relax\def\urlprefix{URL }\fi

\bibitem{STA71}
H.~E. Stanley: {\em Phase transitions and critical phenomena\/} (Clarendon
  Press, Oxford, 1971).

\bibitem{NEW03}
M.~E.~J. Newman: {\em The structure and function of complex networks\/}, SIAM
  Review {\bf 45}, 167 (2003).

\bibitem{BOC18}
S.~Boccaletti, A.~N. Pisarchik, C.~I. del Genio, and A.~Amann: {\em
  Synchronization: {F}rom Coupled Systems to Complex Networks\/} (Cambridge
  University Press, Cambridge, 2018).

\bibitem{HAK83}
H.~Haken: {\em {Synergetics, An Introduction}\/} (Springer, Berlin, 1983), 3rd
  ed.

\bibitem{SCH87}
E.~Sch{\"o}ll: {\em Nonequilibrium Phase Transitions in Semiconductors\/}
  (Springer, Berlin, 1987).

\bibitem{ROD16}
F.~A. Rodrigues, T.~K. D.~M. Peron, P.~Ji, and J.~Kurths: {\em The {Kuramoto}
  model in complex networks\/}, Phys. Rep. {\bf 610}, 1 (2016).

\bibitem{GHO22}
D.~Ghosh, M.~Frasca, A.~Rizzo, S.~Majhi, S.~Rakshit, K.~{Alfaro-Bittner}, and
  S.~Boccaletti: {\em The synchronized dynamics of time-varying networks\/},
  Phys. Rep. {\bf 949}, 1 (2022).

\bibitem{KUR84}
Y.~Kuramoto: {\em Chemical Oscillations, Waves and Turbulence\/}
  (Springer-Verlag, Berlin, 1984).

\bibitem{PIK01}
A.~Pikovsky, M.~Rosenblum, and J.~Kurths: {\em Synchronization: a universal
  concept in nonlinear sciences\/} (Cambridge University Press, Cambridge,
  2001), 1st ed.

\bibitem{PAZ05a}
D.~Paz{\'o}: {\em Thermodynamic limit of the first-order phase transition in
  the kuramoto model\/}, Phys. Rev. E {\bf 72}, 046211 (2005).

\bibitem{GOM11a}
J.~G\'omez-Garde\~nes, S.~G\'omez, A.~Arenas, and Y.~Moreno: {\em Explosive
  synchronization transitions in scale-free networks\/}, Phys. Rev. Lett. {\bf
  106}, 128701 (2011).

\bibitem{BOC16}
S.~Boccaletti, J.~A. Almendral, S.~Guan, I.~Leyva, Z.~Liu,
  I.~Sendi{\~n}a-Nadal, Z.~Wang, and Y.~Zou: {\em Explosive transitions in
  complex networks' structure and dynamics: Percolation and synchronization\/},
  Phys. Rep. {\bf 660} (2016).

\bibitem{SOU19}
R.~M. D'Souza, J.~G\'omez-Garde\~nes, J.~Nagler, and A.~Arenas: {\em Explosive
  phenomena in complex networks\/}, Adv. Phys. {\bf 68}, 123 (2019).

\bibitem{GOM07}
J.~G\'omez-Garde\~nes, Y.~Moreno, and A.~Arenas: {\em Paths to synchronization
  on complex networks\/}, Phys. Rev. Lett. {\bf 98}, 034101 (2007).

\bibitem{ZHA13a}
X.~Zhang, X.~Hu, J.~Kurths, and Z.~Liu: {\em Explosive synchronization in a
  general complex network\/}, Phys. Rev. E {\bf 88}, 010802 (2013).

\bibitem{KUE21}
C.~Kuehn and C.~Bick: {\em A universal route to explosive phenomena\/}, Sci.
  Adv. {\bf 7}, eabe3824 (2021).

\bibitem{YEU99}
M.~K.~S. Yeung and S.~H. Strogatz: {\em Time delay in the {K}uramoto model of
  coupled oscillators\/}, Phys. Rev. Lett. {\bf 82}, 648 (1999).

\bibitem{STR00}
S.~H. Strogatz: {\em From {K}uramoto to {C}rawford: exploring the onset of
  synchronization in populations of coupled oscillators\/}, Physica D {\bf
  143}, 1 (2000).

\bibitem{LEE09}
W.~S. Lee, E.~Ott, and T.~M. Antonsen: {\em Large coupled oscillator systems
  with heterogeneous interaction delays\/}, Phys. Rev. Lett. {\bf 103}, 044101
  (2009).

\bibitem{BRE10h}
M.~Breakspear, S.~Heitmann, and A.~Daffertshofer: {\em Generative models of
  cortical oscillations: neurobiological implications of the {Kuramoto}
  model\/}, Front. Hum. Neurosci. {\bf 4}, 190 (2010).

\bibitem{NAB11}
A.~Nabi and J.~Moehlis: {\em Single input optimal control for globally coupled
  neuron networks\/}, J. Neural Eng. {\bf 8}, 065008 (2011).

\bibitem{BIC20}
C.~Bick, M.~Goodfellow, C.~R. Laing, and E.~A. Martens: {\em Understanding the
  dynamics of biological and neural oscillator networks through exact
  mean-field reductions: a review\/}, J. Math. Neurosci. {\bf 10}, 9 (2020).

\bibitem{OLM14a}
S.~Olmi, A.~Navas, S.~Boccaletti, and A.~Torcini: {\em Hysteretic transitions
  in the {Kuramoto} model with inertia\/}, Phys. Rev. E {\bf 90}, 042905
  (2014).

\bibitem{TUM18}
L.~Tumash, S.~Olmi, and E.~Sch{\"o}ll: {\em {E}ffect of disorder and noise in
  shaping the dynamics of power grids\/}, Europhys. Lett. {\bf 123}, 20001
  (2018).

\bibitem{TUM19}
L.~Tumash, S.~Olmi, and E.~Sch{\"o}ll: {\em {S}tability and control of power
  grids with diluted network topology\/}, Chaos {\bf 29}, 123105 (2019).

\bibitem{TAH19}
H.~Taher, S.~Olmi, and E.~Sch{\"o}ll: {\em Enhancing power grid synchronization
  and stability through time delayed feedback control\/}, Phys. Rev. E {\bf
  100}, 062306 (2019).

\bibitem{HEL20}
F.~Hellmann, P.~Schultz, P.~Jaros, R.~Levchenko, T.~Kapitaniak, J.~Kurths, and
  Y.~Maistrenko: {\em Network-induced multistability through lossy coupling and
  exotic solitary states\/}, Nat. Commun. {\bf 11}, 592 (2020).

\bibitem{JAI01}
S.~Jain and S.~Krishna: {\em A model for the emergence of cooperation,
  interdependence, and structure in evolving networks\/}, Proc. Natl. Acad.
  Sci. {\bf 98}, 543 (2001).

\bibitem{KUE19a}
C.~Kuehn: {\em Multiscale dynamics of an adaptive catalytic network\/}, Math.
  Model. Nat. Phenom. {\bf 14}, 402 (2019).

\bibitem{GRO06b}
T.~Gross, C.~J.~D. D'Lima, and B.~Blasius: {\em Epidemic dynamics on an
  adaptive network\/}, Phys. Rev. Lett. {\bf 96}, 208701 (2006).

\bibitem{PRO05a}
S.~R. Proulx, D.~E.~L. Promislow, and P.~C. Phillips: {\em Network thinking in
  ecology and evolution\/}, Trends Ecol. Evol. {\bf 20}, 345 (2005).

\bibitem{GER96}
W.~Gerstner, R.~Kempter, J.~L. von Hemmen, and H.~Wagner: {\em A neuronal
  learning rule for sub-millisecond temporal coding\/}, Nature {\bf 383}, 76
  (1996).

\bibitem{MEI09a}
C.~Meisel and T.~Gross: {\em Adaptive self-organization in a realistic neural
  network model\/}, Phys. Rev. E {\bf 80}, 061917 (2009).

\bibitem{MIK13}
K.~Mikkelsen, A.~Imparato, and A.~Torcini: {\em Emergence of slow collective
  oscillations in neural networks with spike-timing dependent plasticity\/},
  Phys. Rev. Lett. {\bf 110}, 208101 (2013).

\bibitem{MAR17b}
E.~A. Martens and K.~Klemm: {\em Transitions from trees to cycles in adaptive
  flow networks\/}, Front. Phys. {\bf 5}, 62 (2017).

\bibitem{GRO08a}
T.~Gross and B.~Blasius: {\em Adaptive coevolutionary networks: a review\/}, J.
  R. Soc. Interface {\bf 5}, 259 (2008).

\bibitem{HOR20}
L.~Horstmeyer and C.~Kuehn: {\em Adaptive voter model on simplicial
  complexes\/}, Phys. Rev. E {\bf 101}, 022305 (2020).

\bibitem{GUT11}
R.~Guti{\'e}rrez, A.~Amann, S.~Assenza, J.~G\'omez-Garde\~nes, V.~Latora, and
  S.~Boccaletti: {\em Emerging meso- and macroscales from synchronization of
  adaptive networks\/}, Phys. Rev. Lett. {\bf 107}, 234103 (2011).

\bibitem{KAS17}
D.~V. Kasatkin, S.~Yanchuk, E.~Sch{\"o}ll, and V.~I. Nekorkin: {\em
  {S}elf-organized emergence of multi-layer structure and chimera states in
  dynamical networks with adaptive couplings\/}, Phys. Rev. E {\bf 96}, 062211
  (2017).

\bibitem{BER19}
R.~Berner, E.~Sch{\"o}ll, and S.~Yanchuk: {\em Multiclusters in networks of
  adaptively coupled phase oscillators\/}, SIAM J. Appl. Dyn. Syst. {\bf 18},
  2227 (2019).

\bibitem{BER20}
R.~Berner, J.~Sawicki, and E.~Sch{\"o}ll: {\em Birth and stabilization of phase
  clusters by multiplexing of adaptive networks\/}, Phys. Rev. Lett. {\bf 124},
  088301 (2020).

\bibitem{FEK20}
P.~Feketa, A.~Schaum, and T.~Meurer: {\em Synchronization and multi-cluster
  capabilities of oscillatory networks with adaptive coupling\/}, IEEE Trans.
  Autom. Control {\bf 66}, 3084 (2020).

\bibitem{BER21b}
R.~Berner, S.~Vock, E.~Sch{\"o}ll, and S.~Yanchuk: {\em Desynchronization
  transitions in adaptive networks\/}, Phys. Rev. Lett. {\bf 126}, 028301
  (2021).

\bibitem{LUE16}
L.~L{\"u}cken, O.~V. Popovych, P.~A. Tass, and S.~Yanchuk: {\em
  {N}oise-enhanced coupling between two oscillators with long-term
  plasticity\/}, Phys. Rev. E {\bf 93}, 032210 (2016).

\bibitem{ROE19a}
V.~R{\"o}hr, R.~Berner, E.~L. Lameu, O.~V. Popovych, and S.~Yanchuk: {\em
  Frequency cluster formation and slow oscillations in neural populations with
  plasticity\/}, PLoS ONE {\bf 14}, e0225094 (2019).

\bibitem{SAW21b}
J.~Sawicki, R.~Berner, T.~L{\"o}ser, and E.~Sch{\"o}ll: {\em Modelling tumor
  disease and sepsis by networks of adaptively coupled phase oscillators\/},
  Front. Netw. Physiol. {\bf 1}, 730385 (2022).

\bibitem{BER22}
R.~Berner, J.~Sawicki, M.~Thiele, T.~L{\"o}ser, and E.~Sch{\"o}ll: {\em
  Critical parameters in dynamic network modeling of sepsis\/}, Front. Netw.
  Physiol. {\bf 2}, 904480 (2022).

\bibitem{BER21a}
R.~Berner, S.~Yanchuk, and E.~Sch{\"o}ll: {\em What adaptive neuronal networks
  teach us about power grids\/}, Phys. Rev. E {\bf 103}, 042315 (2021).

\bibitem{THA14}
N.~T.~K. Thanh, N.~Maclean, and S.~Mahiddine: {\em Mechanisms of nucleation and
  growth of nanoparticles in solution\/}, Chem. Rev. {\bf 114}, 7610 (2014).

\bibitem{SCH91}
L.~Schimansky-Geier, C.~Z{\"u}licke, and E.~Sch{\"o}ll: {\em Domain formation
  due to ostwald ripening in bistable systems far from equilibrium\/}, Z. Phys.
  B {\bf 84}, 433 (1991).

\bibitem{AOK09}
T.~Aoki and T.~Aoyagi: {\em Co-evolution of phases and connection strengths in
  a network of phase oscillators\/}, Phys. Rev. Lett. {\bf 102}, 034101 (2009).

\bibitem{MUL11}
L.~Muller, R.~Brette, and B.~Gutkin: {\em Spike-timing dependent plasticity and
  feed-forward input oscillations produce precise and invariant spike
  phase-locking\/}, Front. Comput. Neurosci. {\bf 5}, 45 (2011).

\bibitem{ACE05}
J.~A. Acebr{\'o}n, L.~L. Bonilla, C.~J. P\'{e}rez~Vicente, F.~Ritort, and
  R.~Spigler: {\em The {K}uramoto model: A simple paradigm for synchronization
  phenomena\/}, Rev. Mod. Phys. {\bf 77}, 137 (2005).

\bibitem{Note1}
Note that the choice of $\protect \hat {\omega }$ is arbitrary as it can be
  compensated by the rescaling of $\sigma $, $\epsilon $, and time $t$.

\bibitem{BER19a}
R.~Berner, J.~Fialkowski, D.~V. Kasatkin, V.~I. Nekorkin, S.~Yanchuk, and
  E.~Sch{\"o}ll: {\em Hierarchical frequency clusters in adaptive networks of
  phase oscillators\/}, Chaos {\bf 29}, 103134 (2019).

\bibitem{suppl}
See the supplemental material at ... for details on the numerical implementation, the synchronization transition with equiprobable natural frequencies, with multimodal natural frequencies and for other values of $\beta$, the hysteretic behavior, the antipodal symmetry of the adaptive network model, the derivation and analysis of the mean-field equations and for details on the clustering behavior during the transition to synchrony

\bibitem{Note2}
For the finite time evaluation in the numerical analysis, we use the threshold
  $ \langle \protect \dot {\phi }_i\rangle - \langle \protect \dot {\phi
  }_j\rangle \le 10^{-3}$.

\bibitem{MAI07}
Y.~Maistrenko, B.~Lysyansky, C.~Hauptmann, O.~Burylko, and P.~A. Tass: {\em
  Multistability in the kuramoto model with synaptic plasticity\/}, Phys. Rev.
  E {\bf 75}, 066207 (2007).

\bibitem{GOT15}
G.~A. Gottwald: {\em Model reduction for networks of coupled oscillators\/},
  Chaos {\bf 25}, 053111 (2015).

\bibitem{SMI19}
L.~D. Smith and G.~A. Gottwald: {\em Chaos in networks of coupled oscillators
  with multimodal natural frequency distributions\/}, Chaos {\bf 29}, 093127
  (2020).

\bibitem{HAN18b}
E.~J. Hancock and G.~A. Gottwald: {\em Model reduction for kuramoto models with
  complex topologies\/}, Phys. Rev. E {\bf 98}, 012307 (2018).

\bibitem{SMI20}
L.~D. Smith and G.~A. Gottwald: {\em Model reduction for the collective
  dynamics of globally coupled oscillators: From finite networks to the
  thermodynamic limit\/}, Chaos {\bf 30}, 093107 (2020).

\bibitem{Note3}
Note that the ansatz (\ref {eq:PhiApprox}) is only valid for fully connected
  networks~\cite {HAN18b} but the numerical results show that this
  simplification is justified.

\bibitem{Note4}
Note that the case of an arbitrary number of clusters can be achieved within
  the framework of collective coordinates~\cite {HAN18b,SMI19}.

\bibitem{FEK21}
P.~Feketa, A.~Schaum, and T.~Meurer: {\em {Stability of cluster formations in
  adaptive Kuramoto networks}\/}, IFAC-PapersOnLine {\bf 54}, 14 (2021).

\bibitem{SKA13a}
P.~S. Skardal, D.~Taylor, and J.~G. Restrepo: {\em Complex macroscopic behavior
  in systems of phase oscillators with adaptive coupling\/}, Physica D {\bf
  267}, 27 (2013).

\bibitem{ZHA15a}
X.~Zhang, S.~Boccaletti, S.~Guan, and Z.~Liu: {\em Explosive synchronization in
  adaptive and multilayer networks\/}, Phys. Rev. Lett. {\bf 114}, 038701
  (2015).

\bibitem{AVA18}
V.~Avalos-Gayt\'an, J.~A. Almendral, I.~Leyva, F.~Battiston, V.~Nicosia,
  V.~Latora, and S.~Boccaletti: {\em Emergent explosive synchronization in
  adaptive complex networks\/}, Phys. Rev. E {\bf 97}, 042301 (2018).

\bibitem{PRU10}
H.~R. Pruppacher and J.~D. Klett: {\em Microphysics of Clouds and
  Precipitation\/}, Atmospheric and Oceanographic Sciences Library (Springer,
  Dordrecht, 2010).

\bibitem{MUL01}
J.~W. Mullin: {\em Crystallization\/} (Elsevier, 2001).

\bibitem{OTT08}
E.~Ott and T.~M. Antonsen: {\em Low dimensional behavior of large systems of
  globally coupled oscillators\/}, Chaos {\bf 18}, 037113 (2008).

\bibitem{GKO22}
M.~A. Gkogkas, C.~Kuehn, and C.~Xu: {\em Mean field limits of co-evolutionary
  heterogeneous networks\/}, arXiv:2202.01742   (2022).

\bibitem{TAN97}
H.~Tanaka, A.~J. Lichtenberg, and S.~Oishi: {\em First order phase transition
  resulting from finite inertia in coupled oscillator systems\/}, Phys. Rev.
  Lett. {\bf 78}, 2104 (1997).

\bibitem{JI13}
P.~Ji, T.~K. D.~M. Peron, P.~J. Menck, F.~A. Rodrigues, and J.~Kurths: {\em
  Cluster explosive synchronization in complex networks\/}, Phys. Rev. Lett.
  {\bf 110}, 218701 (2013).

\end{thebibliography}
\end{document}